\documentclass[fleqn,usenatbib,useAMS]{mnras}
\usepackage{lmodern}
\usepackage{cite}
\usepackage[dvips]{graphicx}
\usepackage{wrapfig}
\graphicspath{{image/}}
\usepackage{color}
\usepackage{amsmath}
\usepackage{amssymb}
\usepackage{rotating}
\usepackage{appendix}
\usepackage{ulem}
\usepackage{soul, xcolor}

\newcommand{\HI}{\ion{H}{I}}

\newcommand{\TB}{\delta T_{\rm b}}
\newcommand{\MSUN}{{\rm M}_{\odot}}

\newcommand{\XHI}{x_{\ion{H}{I}}}
\newcommand{\XHII}{x_{\ion{H}{II}}}
\newcommand{\TS}{T_{\rm s}}
\newcommand{\TK}{T_{\rm K}}
\newcommand{\TCMB}{T_{\gamma}}

\newcommand{\OmegaB}{\Omega_{\rm B}}
\newcommand{\Omegam}{\Omega_{\rm m}}
\def\xb{\bar{x}_{\ion{H}{I}}}
\def\kk{\mathbfit{k}}
\def\U{\mathbfit{U}}
\def\dtb{\delta T_{\rm b}}
\def\ttb{\delta \tilde{T}_{{\rm b}2}}
\def\tvec{\boldsymbol{\theta}}
\def\cl{\mathcal{C}_{\ell}}
\newcommand{\CommentOut}[1]{}

\title[Cosmic Dawn 21-cm MAPS]{Studying the Multi-frequency Angular Power Spectrum of the Cosmic Dawn 21-cm Signal}
\author[A. K. Shaw et al.]{Abinash Kumar Shaw$^1$\thanks{E-mail: \href{mailto:abinashkumarshaw@gmail.com}{abinashkumarshaw@gmail.com}}, Raghunath Ghara$^1$, Saleem Zaroubi$^{2,3}$, Rajesh Mondal$^{4,5}$, Garrelt Mellema$^5$,
\newauthor Florent Mertens$^{3,6}$, L{\'e}on V. E. Koopmans$^3$, Beno{\^i}t Semelin$^6$\\
$^1$Astrophysics Research Center of the Open University (ARCO), The Open University of Israel, 1 University Road, Ra'anana 4353701, Israel.\\
$^2$Department of Natural Sciences, The Open University of Israel, 1 University Road, Ra'anana 4353701, Israel.\\
$^3$Kapteyn Astronomical Institute, University of Groningen, PO Box 800, 9700AV Groningen, The Netherlands.\\
$^4$Department of Astrophysics, School of Physics and Astronomy, Tel Aviv University, Tel Aviv 69978, Israel.\\
$^5$Department of Astronomy \& Oskar Klein Centre, Stockholm University, AlbaNova, SE-10691 Stockholm, Sweden.\\
$^6$LERMA, Observatoire de Paris, PSL Research University, CNRS, Sorbonne Universit{\'e}, F-75014 Paris, France.}

\date{\today}
\pubyear{2023}

\begin{document}
\setstcolor{red} % setting strike-through color red
\label{firstpage}
\pagerange{\pageref{firstpage}--\pageref{lastpage}}
\maketitle

%--------------------------------------------------------------------

\begin{abstract}
The light-cone (LC) anisotropy arises due to cosmic evolution of the cosmic dawn 21-cm signal along the line-of-sight (LoS) axis of the observation volume. The LC effect makes the signal statistically non-ergodic along the LoS axis. The multi-frequency angular power spectrum (MAPS) provides an unbiased alternative to the popular 3D power spectrum as it does not assume statistical ergodicity along every direction in the signal volume. Unlike the 3D power spectrum which mixes the cosmic evolution of the 21-cm signal along the LoS $k$ modes, MAPS keeps the evolution information disentangled. Here we first study the impact of different underlying physical processes during cosmic dawn on the behaviour of the 21-cm MAPS using simulations of various different scenarios and models. We also make error predictions in 21-cm MAPS measurements considering only the system noise and cosmic variance for mock observations of HERA, NenuFAR and SKA-Low. We find that $100~{\rm h}$ of HERA observations will be able to measure 21-cm MAPS at $\geq 3\sigma$ for $\ell \lesssim 1000$ with $0.1\,{\rm MHz}$ channel-width. The better sensitivity of SKA-Low allows reaching this sensitivity up to $\ell \lesssim 3000$. Note that due to the difference in the frequency coverage of the various experiements, the CD-EoR model considered for NenuFAR is different than those used for the HERA and SKA-Low predictions. Considering NenuFAR with the new model, measurements $\geq 2\sigma$ are possible only for $\ell \lesssim 600$ with $0.2\,{\rm MHz}$ channel-width and for a ten times longer observation time of $t_{\rm obs} = 1000~{\rm h}$. However, for the range $300 \lesssim \ell \lesssim 600$ and $t_{\rm obs}=1000~{\rm h}$ more than $3\sigma$ measurements are still possible for NenuFAR when combining consecutive frequency channels within a $5 ~{\rm MHz}$ band.
\end{abstract}

%--------------------------------------------------------------------


%--------------------------------------------------------------------

\section{Introduction}
\label{sec:intro}
The study of Cosmic Dawn (CD) and Epoch of Reionization (EoR) is very crucial to understand the first luminous objects in our Universe and how they caused the last important phase change in the inter-galactic medium (IGM) on cosmological scales. X-ray emission from these first objects (stars, quasars etc.) are expected to heat up the cold neutral hydrogen atom (\HI) during the CD which is then followed by UV photo-ionization of \HI~during the EoR \citep[see e.g.][for reviews]{Pritchard_2006, Zaroubi_2013, Shaw_review}. Unfortunately, our current understanding of the IGM during the EoR is limited to only a few indirect observations such as the measurements of Thomson scattering optical depth of the cosmic microwave background (CMB) photons with the free electrons \citep[e.g.][]{Planck_Cosmo_2018} and the presence of complete Gunn-Peterson troughs in high-redshift quasar spectra \citep[e.g.][]{becker_2001, Fan_2006,gallerani_2006, Becker_2015}. Furthermore, results from the observations of high-redshift Ly-$\alpha$ emitters \citep[e.g.][]{Hu_2010, Ota_2017, Ishigaki_2018, Morales_2021}, the Ly-$\alpha$ damping wings in high-redshift quasar spectra \citep[e.g.][]{Banados_2018, Davies_2018, Wang_2020, Durovcikova_2020, Reiman_2020, Yang_2020, Greig_2022}, Lyman break galaxies \citep{Mason_2018, Hoag_2019, Naidu_2020}, the measurements of the effective optical depth of the Ly-$\alpha$ (and Ly-$\beta$) forests \citep[e.g.][]{McGreer_2014, Kulkarni_2019, Raste_2021, Zhu_2021, Zhu_2022} and the high-$z$ Ly-$\alpha$ emitters clustering masurements \citep[e.g.][]{Faisst_2014, santos_Lya-2016, Sobacchi_2015, Wold_2022} have put some loose bounds on the average \HI~fraction in the IGM. These indirect observations in combination suggest a bound on the timing of reionization to be between $z\approx 5.5-12$ \citep[e.g.][]{Robertson_2015, Mitra_2018, Dai_2019, Qin_2021}. Besides the ionization state, there have been attempts to constrain IGM temperature and photoionization rates using Ly-$\alpha$ forests observations \citep[][]{Gaikwad_2020}. However these constraints are not very tight and a direct probe of the IGM is required to answer several fundamental questions regarding CD-EoR such as what is the exact duration of these epochs, how the heating and reionization have progressed, what were the sources involved and what were their properties.

The most promising direct probe to the CD-EoR is the redshifted 21-cm radiation which is emitted due to hyperfine transition of \HI~in the IGM \citep[e.g.][]{Wouthuysen_1952, Field_1958, sunyaev_1972, Hogan_1979, scott_1990, Bharadwaj2001}. Concerted efforts have been put into observing the brightness temperature fluctuations of the redshifted 21-cm signal using low frequency radio-interferometers. Several current telescopes, such as LOFAR\footnote{\url{https://www.astron.nl/telescopes/lofar}\label{ft:lofar}} \citep{vanHaarlem_2013}, PAPER\footnote{\url{http://eor.berkeley.edu}} \citep{Parsons_2010}, NenuFAR\footnote{\url{https://nenufar.obs-nancay.fr/en/homepage-en}\label{ft:nenu}} \citep{Zarka_2018}, uGMRT\footnote{\url{http://www.gmrt.ncra.tifr.res.in}} \citep{Swarup_1991, Gupta_2017}, MWA\footnote{\url{https://www.mwatelescope.org}} \citep{Tingay_2013}, HERA\footnote{\url{https://reionization.org}\label{ft:hera}} \citep{DeBoer_2017} are trying to measure the redshifted 21-cm signal statistically.

Despite these efforts, the 21-cm signal has not yet been measured due to several observational challenges. The cosmological 21-cm signal is $10^4-10^5$ times weaker than the galactic and extra-galactic foregrounds \citep[e.g.][]{Ali_2008, Bernardi_2009, Bernardi_2010, Abhik_2012, Beardsley_2016}. In addition, the system noise, systematics and calibration errors and the human-made radio-frequency interference poses a challenge to the direct detection of the CD-EoR 21-cm signal . However, recent improvements in the foreground mitigation methods \citep[e.g.][]{Datta_2010, Mertens_2018, Hothi_2021} and calibration techniques \citep[e.g.][]{Kern_2019, Mevius_2021, Gan_2023} have led us to useful upper limits on spherically-averaged 3D power spectrum (e.g. LOFAR: \citealt{Patil_2017, Gehlot_2019, LOFAR_Mertens_2020}; GMRT: \citealt{Paciga_2011, Paciga_2013}; MWA: \citealt{Li_2019, Trott2020}; PAPER: \citealt{Cheng_2018, Kolopanis_2019}; HERA: \citealt{Abdurashidova_2022a, Abdurashidova_2022b}; OVRO-LWA: \citealt{Eastwood_2019}). Despite these challenges, the upper limits are constantly improving and a few recent upper limits have started ruling out extreme models of the high-$z$ IGM (e.g. LOFAR: \citealt[][]{Ghara_2020, Greig_2020b, Mondal_2020b}; MWA: \citealt[][]{Greig_2020a, Ghara_2021}; HERA: \citealt[][]{Abdurashidova_2022a, Abdurashidova_2022b}). The upcoming SKA-Low\footnote{\url{https://www.skao.int}\label{ft:ska}} is expected to measure the 21-cm 3D power spectrum from both CD and EoR within about a hundred hours of observing time \citep[e.g.][]{koopmans2014, shaw_2019}. SKA-Low will also be able to produce images of 21-cm signals during CD \citep[][]{Ghara_2016} and EoR \citep[][]{Mellema_2013} within a thousand hours of observing time.

These radio-interferometric observations record the 3D distribution of the \HI~21-cm signal, where the third direction in the data volume is constructed from several channels across the frequency bandwidth. The 21-cm signal measured at different frequencies corresponds to the signal originating at different redshifts and thereby at different cosmic times. Thus, the imprint of cosmic evolution in the data makes the observed 21-cm signal statistically non-ergodic along the frequency axis. This effect, also known as the `light-cone (LC) effect' is inevitable in any line-intensity mapping observations. \citet{Barkana_2006} were the first to analytically quantify the line-of-sight (LoS) anisotropy introduced in the EoR 21-cm signal due to the LC effect. The LC effect also biases the statistical estimators which assume statistical ergodicity and periodicity in the data in all directions such as 3D \textit{N}-point correlation functions and their Fourier conjugates. Several studies have quantified the LC effect on the two-point correlations \citep[][]{Zawada_2014} and the 3D power spectrum \citep[e.g.][]{Light_cone_I, La_Plante_2014, Light_Cone_II} using the numerically simulated EoR 21-cm signal. \citet{Light_cone_I} found that the effect is about $50~\%$ on scales corresponding to $k=0.08~{\rm Mpc}^{-1}$ along the LoS. \citet{Light_Cone_II} found that the LC effect shows dramatic changes in 3D power spectrum only during the initial ($\xb \approx 0.2$) and final ($\xb \approx 0.8$) stages of their fiducial reionization model. However, they also argue that impact would be more important for shorter reionization histories. \citet{Ghara_2015b} are the first to study the effect of LC anisotropy on CD 21-cm 3D power spectra. They concluded that the LC effects are important for the CD 21-cm signal which is mostly governed by the spin temperature fluctuations that in turn are controlled by inhomogeneous X-ray heating and the Ly-$\alpha$ coupling of the IGM. They reported suppression(enhancement) in the peaks(dips) by factors $\geq 0.6(2.0)$ for their fiducial model at large-scales ($k\approx 0.05\,{\rm Mpc}^{-1}$) using simulated LC boxes having volume $V = [200h^{-1}~{\rm Mpc}]^3$. Unlike during the EoR, the LC bias becomes dramatically important during CD, even at small-scales ($k\sim 1\,{\rm Mpc}^{-1}$).

Recently, \citet{Mondal_2017} have proposed the use of Multi-frequency Angular Power Spectrum (MAPS) to evade the issue of the LC effect in the EoR 21-cm signal while quantifying its two-point statistics. Unlike the 3D power spectrum, MAPS is a non-stationary estimator along the LoS axis that captures the two-point statistics without any bias originating due to LC anisotropy in the observed signal. MAPS also has potential to recover the evolution of EoR \citep{Mondal_2019a}. Later, \citet{Mondal_2020a} have quantified the EoR 21-cm MAPS and predicted its detectability in context of future SKA-Low observations. Recently \citet{Trott_2022} have produced first upper limits on the EoR 21-cm MAPS within $z=6.2-7.5$ using the MWA data. MAPS statistics is expected to contain more unbiased information of the 21-cm signal than 3D power spectrum. This extra information can be exploited to put more stringent constraints over the models of CD-EoR \citep{Mondal_2022}. Along a similar line, this work focuses on estimating the CD 21-cm MAPS using a suite of CD-EoR 21-cm simulations. We aim to predict the detectability of CD 21-cm MAPS using several current (HERA, NenuFAR) and upcoming (SKA-Low) telescopes. Unlike recent works which were limited to estimating the EoR 21-cm MAPS within a small frequency bandwidth (LoS depth $\lesssim 1~{\rm Gpc}$), here we have computed 21-cm MAPS over a large frequency bandwidth (LoS depth $\approx 2.1~{\rm Gpc}$; redshift range $z\approx 6-19$) that comprises both the CD and EoR.

The organization of this paper is as follows. Section \ref{sec:method} briefly describes the simulation methodology, the impact of LC effect on CD-EoR 21-cm 3D power spectrum and the formulation and implementation of the MAPS statistics. We then discuss the theoretical implications of different physical processes and source models on 21-cm MAPS in Section \ref{sec:astro}. In Section \ref{sec:detect}, we discuss the detectability of the CD-EoR 21-cm MAPS considering HERA, NenuFAR and SKA-Low telescopes. Finally, we summarize our work and conclude in Section \ref{sec:summ}. This work uses the cosmological parameter values $\Omegam=0.27$, $\Omega_{\Lambda}=0.73$, $\OmegaB=0.044$ and $h=0.7$ taken from \citet{Hinshaw_2013}. These parameters have been used in the \textit{N}-body simulation for this work.
%%%%%%%%%%%%%%%%%%%%%%%%%%%%%%%%%%%%%%%%%%%%%%%%%%%%%%%%%%%%%%%%%%%%%%%%%%%%%%%%

\begin{figure*}
\begin{center}
\includegraphics[scale=0.5]{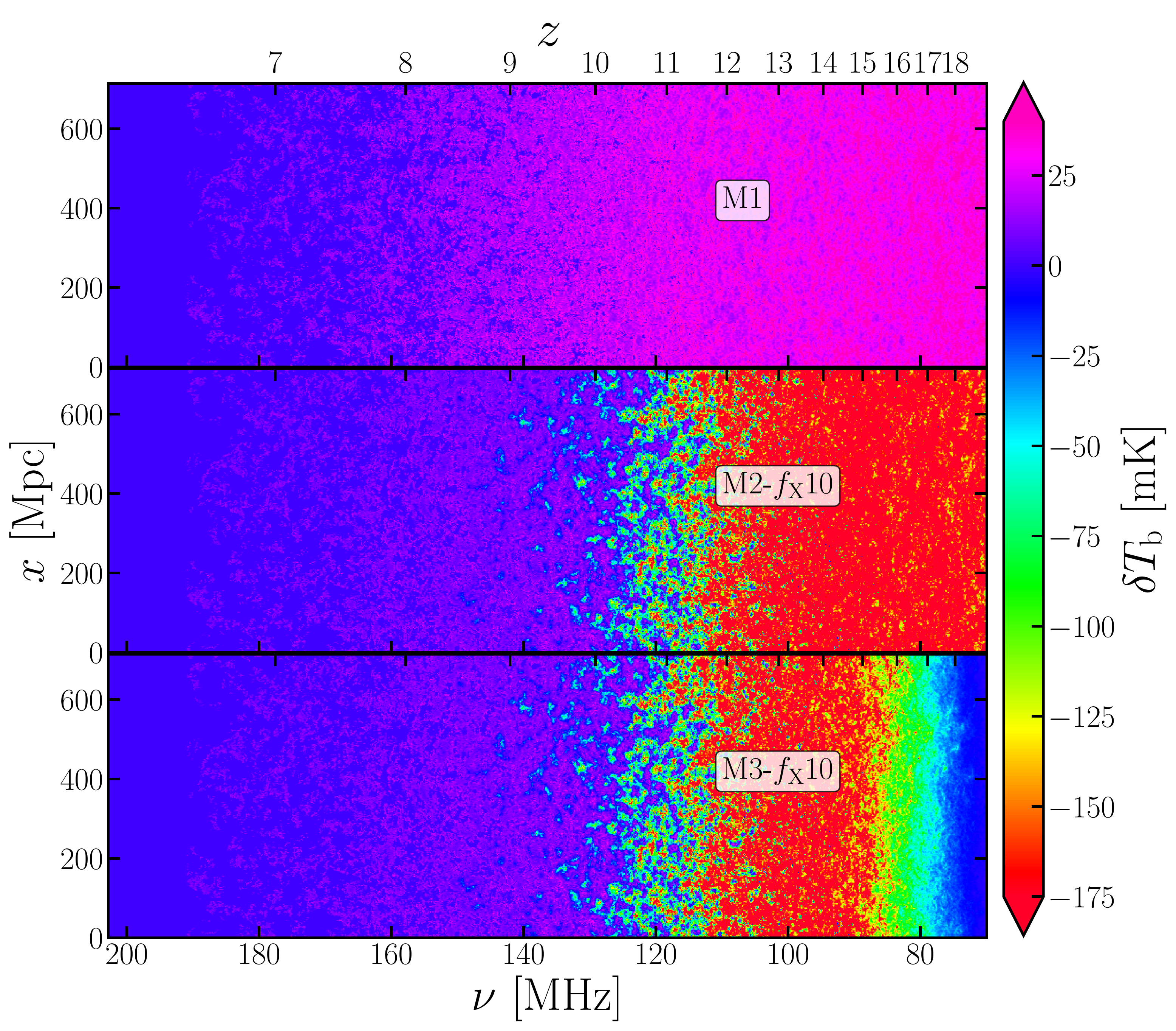}
    \caption{Light-cone slices of simulated CD-EoR 21-cm signal as a function of observing frequency(redshift). The three panels corresponds to three different models corresponding to a heating scenario with $f_{\rm X}=10$. The details are mentioned in section \ref{sec:sim}.}
   \label{fig:tb_slice_model}
\end{center}
\end{figure*}

\begin{figure*}
\begin{center}
\includegraphics[scale=0.5]{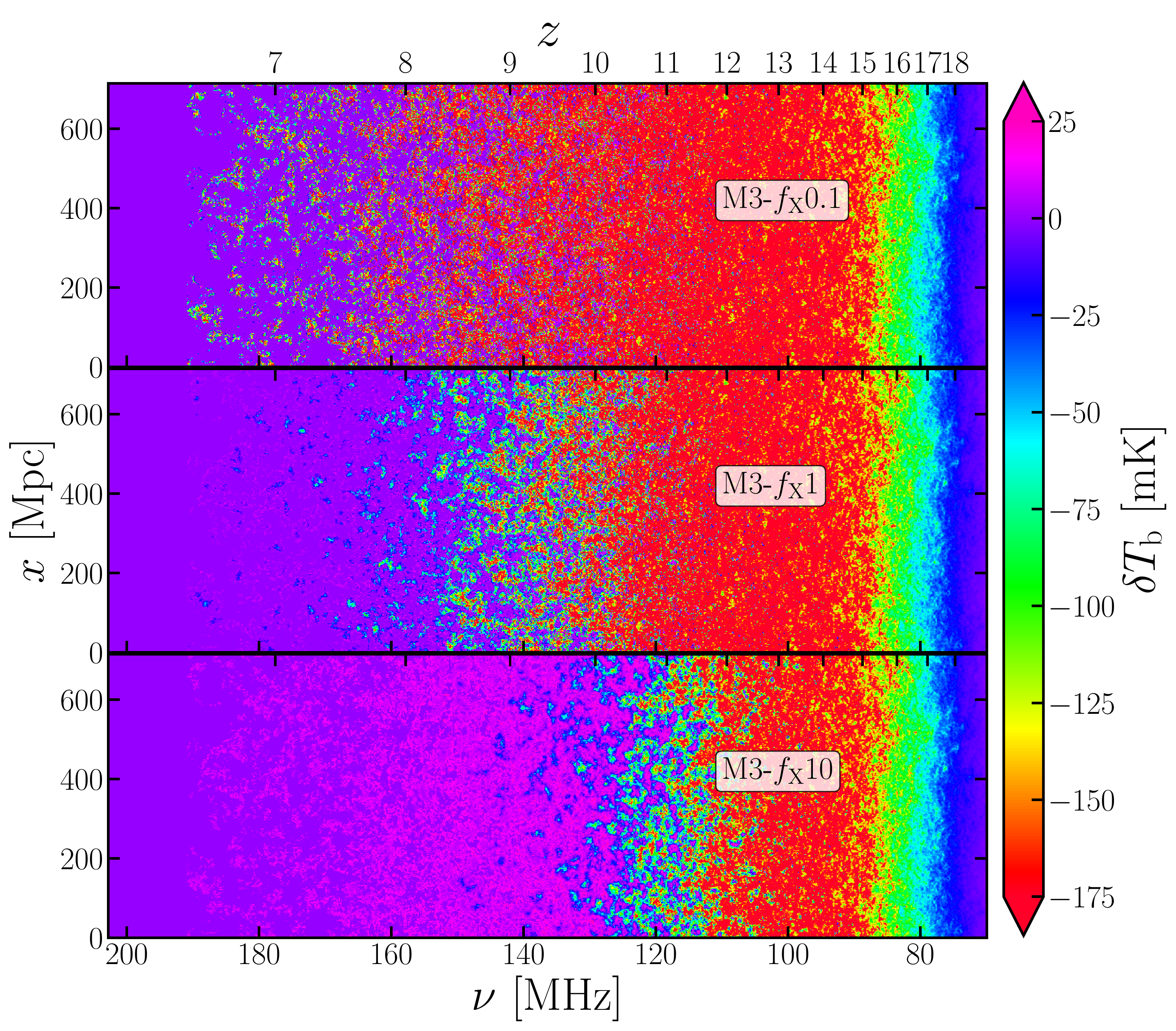}
    \caption{Light-cone slices of simulated CD-EoR 21-cm signal as a function of observing frequency(redshift). The three panels corresponds to three different heating scenarios corresponding to model M$3$. The details are mentioned in section \ref{sec:sim}.}
   \label{fig:tb_slice_scenario}
\end{center}
\end{figure*}

\section{Methodology}\label{sec:method}
We first briefly describe our LC simulations in this section. Next, we study the impact of the LC effect on the 3D power spectrum estimator using our LC simulation. Finally, we describe the machinery of multifrequency angular power spectrum after motivating it as an alternative to the 3D power spectrum.
\subsection{Simulating the 21-cm signal Light-cone}
\label{sec:sim}
We use the {\sc grizzly} code \citep{Ghara_2015a, raghunath_RT} to simulate 21-cm signal coeval cubes at different redshifts during the CD and EoR. This code is based on a one-dimensional radiative transfer method which takes the cosmological density and velocity fields and the halo catalogues as input to simulate the differential brightness temperature ($\TB$) coeval cubes of the 21-cm signal for a source model. Later, we use these coeval $\TB$ cubes to generate a $\TB$ light-cone that spans the CD and EoR. Here, we briefly describe the steps employed to generate the 21-cm signal light-cones which we will use in this study.

The cosmological density and velocity fields and the halo catalogues, used in this study, are taken from results of the PRACE\footnote{Partnership for Advanced Computing in Europe: \url{http://www.prace-ri.eu/}} project PRACE4LOFAR. These inputs are from a dark-matter only \textit{N}-body simulation that was run using the code {\sc cubep}$^3${\sc m}\footnote{\url{http://wiki.cita.utoronto.ca/mediawiki/index.php/CubePM}} \citep{Harnois_2012}. Here, we use gridded density and velocity fields in 3D comoving cubes with length $500~h^{-1}$~ comoving megaparsec (Mpc) \citep[see e.g,][]{Dixon_2015, Giri_2019a}. The volume is gridded into $300^3$ cubical voxels. We use $59$ snapshots in the redshift range $20 \gtrsim z \gtrsim 6.5$ in equal time steps of $11.4$ Myr. The dark matter halos were identified using a spherical overdensity halo finder \citep{Watson_2013}. The minimum mass of the dark matter halos is $\sim 10^9\,{\rm M}_\odot$, having at least $\approx 25$ dark matter particles.

The source model used in this study assumes that each dark matter halo with mass larger than $10^9 ~\MSUN$ hosts sources which emit UV as well as X-rays. In particular, we consider a combination of galaxy and mini-quasar type sources. We assume that the stellar mass of a galaxy ($M_\star$) is linearly proportional to the host dark matter halo mass $M_{\rm halo}$, \textit{i.e.}, $M_\star=f_\star \left(\frac{\OmegaB}{\Omegam}\right) M_{\rm halo}$. The quantity $f_\star$ is the fraction of baryons residing in stars within a galaxy. We fix $f_\star=0.02$ in this study \citep{Behroozi_2015, Sun_2016}. {\sc grizzly} also uses a galaxy spectral energy distribution (SED) per stellar mass in the 1D radiative transfer step. The galaxy SED per stellar mass is generated for standard star formation scenarios using the code {\sc pegase2}\footnote{\href{http://www2.iap.fr/pegase/}{http://www2.iap.fr/pegase/}} \citep{Fioc_1997}. The initial mass function (IMF) of the stars, within the mass range $1-100 ~\MSUN$ in the galaxies, is assumed as a Salpeter IMF and the galaxy metallicity is assumed to be $0.1~Z_\odot$ throughout the EoR. The emission rate of ionizing photons per unit stellar mass $\dot N_i$ from a halo is controlled by the ionization efficiency ($\zeta$) parameter used in {\sc grizzly}. $\dot N_i=\zeta\times 2.85\times 10^{45}  ~{\rm s^{-1}} ~\MSUN^{-1}$ where $\zeta=1$ corresponds to $\dot N_i$ of {\sc pegase2} SED per stellar mass. We fix $\zeta=0.1$ for all the reionization models considered in this study. This results in EoR ending around $z\approx 6.5$. On the other hand, the emission rate of X-ray photons per unit stellar mass $\dot N_{\rm X}$ from a halo is controlled by X-ray heating efficiency parameter $f_{\rm X}$, where $\dot N_{\rm X} = f_{\rm X} \times  10^{42} ~\rm s^{-1} ~\MSUN^{-1}$. The value of $\dot N_{\rm X}$ for $f_{\rm X}=1$ is consistent with the measurements of $0.5-8~{\rm keV}$ band HMXBs SEDs in local star-forming galaxies \citep{Mineo_2012}. The X-ray SED of the source is modelled using a power-law of energy $E$ as $I_{\rm X}(E) \propto E^{-\alpha}$. We fix the spectral index $\alpha=1.2$ throughout the work \citep{Vanden_Berk_2001, Vignali_2003}. The X-ray band here spans from $0.1$ to $10$ keV. We vary $f_{\rm X}$ parameters to model different heating (or Cosmic Dawn) scenarios of the IGM. The details of the SEDs can be found in \citet{Ghara_2015a} and \citet{Islam_2019}.

Subsequently, we create a library of a large number of one-dimensional profiles of hydrogen ionization fraction $\XHII$, and gas temperature $\TK$ around isolated sources for different combinations of stellar masses, redshifts and density contrast. We use these profiles to generate coeval cubes of neutral fraction $\XHI$ and $\TK$ at different redshifts. The 21-cm signal strength also crucially depends on the Ly-$\alpha$ photons flux. The Ly-$\alpha$ flux maps are generated assuming a $1/R^2$ decrease of the Ly-$\alpha$ photons flux with radial distance $R$ from the source. We refer the readers to \citet{Ghara_2015a} and \citet{Islam_2019} for more details about these calculations. The $\TK$ cubes and the Ly-$\alpha$ flux maps are used to generate the spin temperature $T_{\rm s}$ maps assuming that the collisional coupling is negligible at the concerned redshifts. These $\XHI$, $T_{\rm s}$ cubes and the density fields are then used to generate the brightness temperature maps of the 21-cm signal \citep[see e.g.,][]{madau_1997, Furlanetto2006}.
\begin{equation}
    \begin{split}
        \TB (\mathbfit{n}, \nu)  = 27 & ~  \XHI (\mathbfit{n}, z) \left(1+\delta_{\rm B}(\mathbfit{n}, z)\right) \left(\frac{\OmegaB h^2}{0.023}\right)\\
&\times \left(\frac{0.15}{\Omegam h^2}\frac{1+z}{10}\right)^{1/2}\left(1-\frac{\TCMB(z)}{T_{\rm s}(\mathbfit{n}, z)}\right)\,\rm{mK}~,
    \end{split}
    \label{eq:brightnessT}
\end{equation}
where $\mathbfit{n}$ is the position of the observed region in the sky and $\nu=1420.406/(1+z)~{\rm MHz}$ is the observation frequency. The CMB brightness temperature is $\TCMB(z) = 2.725 \times (1+z)~{\rm K}$. We then incorporate the effect due to the peculiar velocity of the gas using the cell movement method described by \citet{Mao_2012}. Finally, we used these coeval cubes of the 21-cm signal brightness temperature to create the LC which accounts for the evolution of the 21-cm signal with redshift. The detailed method to implement the LC effect can be found in \citet{Ghara_2015b}.

We choose $f_{\rm X}=10$ for our fiducial model as it produces heating and reionization peaks in 21-cm 3D power spectra (see Fig. \ref{fig:3dps}) within the redshift range considered here that is comparable to the earlier studies \citep[e.g.][]{Mesinger_2013}. The CD for this model starts with a weak signal due to weak Ly-$\alpha$ background around $z\sim 20$ (see the bottom panel of Fig. \ref{fig:tb_slice_model}). With the Ly-$\alpha$ coupling becoming efficient at $z\lesssim 13$, X-ray heating dominates the fluctuations of the signal at $9\lesssim z \lesssim 15$. At $z\lesssim 9$, the signal fluctuation is dominated by ionization fluctuation. 
\begin{table}
\begin{center}
\small
\tabcolsep 12pt
\renewcommand\arraystretch{1.5}
\caption{An overview of the different CD-EoR models and heating scenarios considered in this paper.}
%Tabulates an overview of the different models and heating scenarios considered in this paper. We consider all the dark matter halos with mass larger than $10^9 ~\MSUN$ contribute to reionization, heating and Ly-$\alpha$ coupling. The ionization efficiency parameter $\zeta$ is fixed to $0.1$. M$3$-$f_{\rm X}10$ is our fiducial simulation.
\begin{tabular}{c c c l}
\hline
Models & $f_{\rm X}$ & $\TS$ & 	 Notations \\
\hline
\hline
Model $1$ &  --   & $\TS>>\TCMB$ &   M$1$ 	\\
Model $2$ &  $10$  &  $\TS=\TK$ & M$2$-$f_{\rm X}10$	\\
Model $3$ & $10$ & Self-Consistent & M$3$-$f_{\rm X}10$	\\
Model $3$ & $1$ & Self-Consistent & M$3$-$f_{\rm X}1$	\\
Model $3$ & $0.1$ & Self-Consistent & M$3$-$f_{\rm X}0.1$ \\
\hline
\end{tabular}
\end{center}
{$^{\ddagger \ddagger}$ The models above consider all the dark matter halos with mass larger than $10^9 ~\MSUN$ contribute to reionization, heating and Ly-$\alpha$ coupling. The ionization efficiency parameter $\zeta$ is fixed to~$0.1$. M$3$-$f_{\rm X}10$ is our fiducial simulation.}
\label{tab:sim}
\end{table}

The most important physical processes that determine the strength and fluctuations of the simulated 21-cm signal are the Ly-$\alpha$ coupling, X-ray heating and the ionization due to UV photons. To study the impact of these processes separately, we consider three different models which are shown in Fig. \ref{fig:tb_slice_model}.
\begin{itemize}
    \item Model $1$: Considers saturated X-ray heating and  Ly-$\alpha$ coupling from the very beginning of the CD (here $z>20$). In other words, we assume $\TS >> \TCMB$. Thus the fluctuations in the signal are determined by the density fluctuations (important during the early stages of reionization) and ionization fluctuations (important during the later stages of reionization). We label this model as `M$1$'.
    \item  Model $2$: Considers X-ray heating self-consistently while assuming the Ly-$\alpha$ coupling is strong from the beginning of the CD. This sets $\TS=\TK$ and $\dtb$ is predominantly governed by matter fluctuations initially. As soon as X-rays start heating the IGM, $\dtb$ starts being governed by the fluctuations in $\TS$. Thus, we will be able to distinguish the impact of the X-ray heating on the 21-cm MAPS by comparing this model with M$1$. In this simulation we set $f_{\rm X}=10$. We denote this model by `M$2$-$f_{\rm X}10$' in our following discussions.
    \item Model $3$: The standard scenario with self-consistent Ly-$\alpha$ coupling and X-ray heating with $f_{\rm X}=10$. This is our fiducial model which we denote by `M$3$-$f_{\rm X}10$'.
\end{itemize}

We also study the impact of different X-ray heating scenarios on the CD 21-cm MAPS. Thus, for Model $3$, we consider two other heating scenarios correspond to $f_{\rm X}=1$ (M$3$-$f_{\rm X}1$) and $f_{\rm X}=0.1$ (M$3$-$f_{\rm X}0.1$). Fig. \ref{fig:tb_slice_scenario} shows the light-cones of these two heating scenarios in addition to our fiducial model M$3$-$f_{\rm X}10$. As expected, we note that the X-ray heating of IGM is delayed for a smaller value of $f_{\rm X}$. In fact, the fraction of emission regions for $f_{\rm X}=0.1$ is negligible. Table \ref{tab:sim} lists all the different CD source models and X-ray heating scenarios that have been considered in this work.

%-----------------------------------------------------------------------------------------------------------%
%%%%%%%%%%%%%%%%%%%%%%%%%%%%%%%%%%%%%%%%%%%%%%%%%%%%%%%%%%%%%%%%%%%%%%%%%%
\subsection{Light-cone effect on 3D power spectrum}\label{sec:3dps}

\begin{figure}
	\begin{center}
		\includegraphics[scale=0.32]{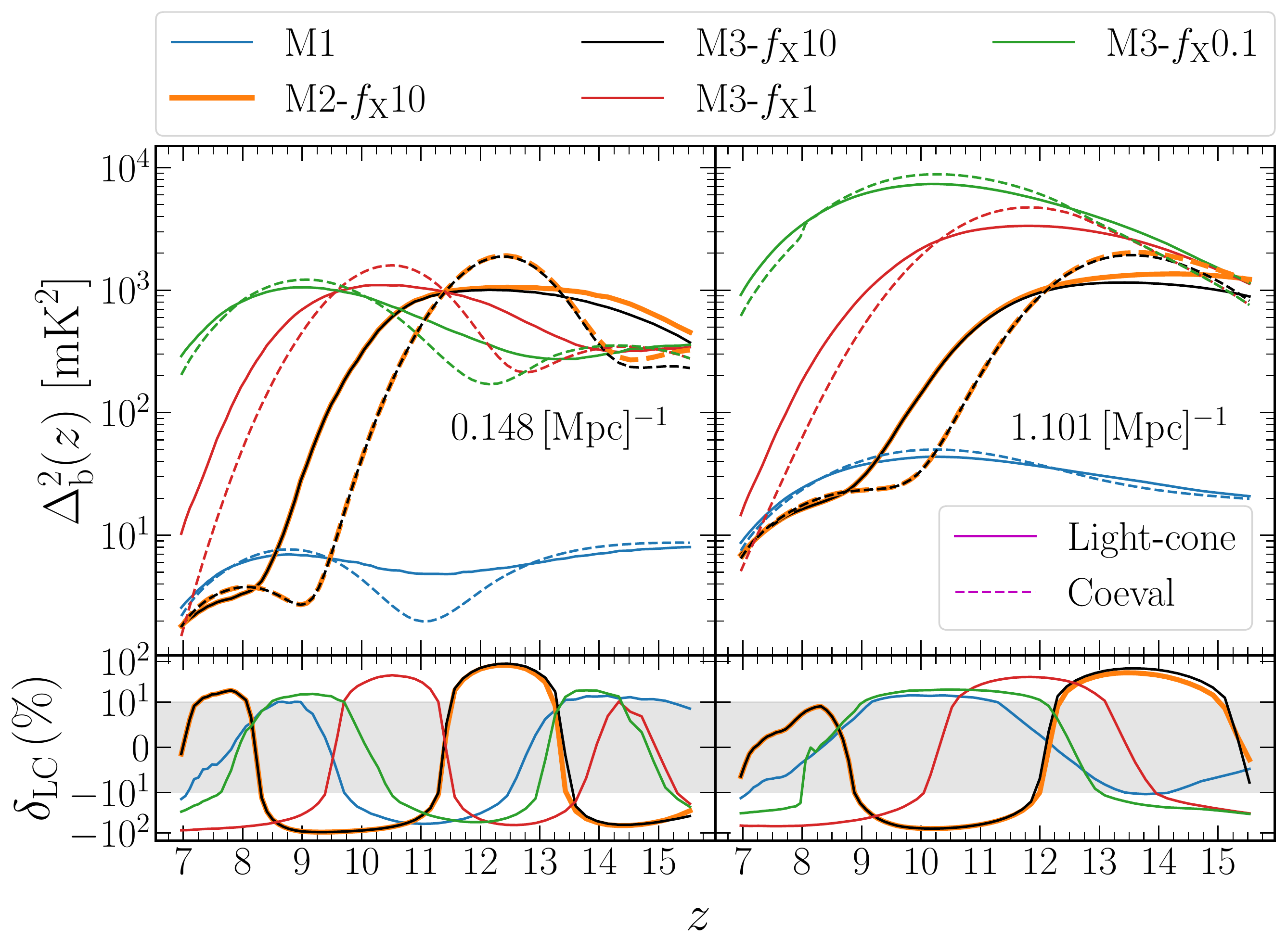}
		\caption{Dimensionless 3D power spectrum $\Delta^2_{\rm b}(k,z)=k^3P(k,z)/(2\pi^2)$ computed using LC and coeval signals as a function of redshift $z$. Two columns are for two different $k$ modes as written in the panels. Five colors here correspond to the five different simulations (see Table \ref{tab:sim}). Solid lines in the top sub-panels represent 3D power spectra estimated from Light-cone boxes whereas the dashed lines represent those computed using coeval cubes. Solid lines in the bottom sub-panels represents percentage deviations in the 3D power spectrum arising due to LC anisotropy. The gray shaded region demarcates the linear scale on the symmetric log axis.}
		\label{fig:3dps}
	\end{center}
\end{figure}

The 3D power spectrum of the CD-EoR 21-cm signal at any wavenumber $\kk$ can be written as
\begin{equation}
    P(\kk, z) = V^{-1} \langle \delta \tilde{T}_{\rm b}(\kk, z) \, \delta \tilde{T}_{\rm b}(-\kk, z) \rangle~,
    \label{eq:3dpk}
\end{equation}
where $V$ is the signal volume centered at redshift $z$ and $\langle \cdots \rangle$ denotes an ensemble average. Here $\delta \tilde{T}_{\rm b}(\kk, z)$ is the Fourier transformation of the 3D signal $\dtb(\mathbfit{n}, z)$ centered at redshift $z$ which inherently assumes statistical ergodicity and periodicity in the signal volume along all directions. Using eq. (\ref{eq:3dpk}), we estimate 3D power spectrum of the LC 21-cm signal averaged over $\kk$ modes in logarithmically-spaced spherical bins. The solid lines in the top sub-panels of Fig. \ref{fig:3dps} show the dimensionless 3D power spectrum $\Delta^2_{\rm b}(z)=k^3P(z)/(2\pi^2)$ computed from the LC boxes shown in Figs. \ref{fig:tb_slice_model} and \ref{fig:tb_slice_scenario}. Here we use a cubical box of volume $V=[500 h^{-1}~{\rm Mpc}]^3$, centered at desired redshifts (frequency channels) within the LC cube, to obtain $\Delta^2_{\rm b, LC}(k,z)$. We also estimate $\Delta^2_{\rm b,CC}(k,z)$ from coeval cubes of similar volume simulated at redshifts the same as the chosen central redshifts in the LC signal. The evolution of $\Delta^2_{\rm b, LC}$ (solid lines) and $\Delta^2_{\rm b, CC}$ (dashed lines) as a function of $z$ are shown in the top sub-panels of Fig. \ref{fig:3dps} for comparison. We present the results for large ($k=0.148~{\rm Mpc}^{-1}$) and small ($k=1.101~{\rm Mpc}^{-1}$) length-scales, respectively, in the left and right panels. It is important to note that the LC effect tries to flatten the peaks and troughs in the 3D power spectrum due to smoothing as already demonstrated in previous studies \citep[e.g.,][]{Light_cone_I, La_Plante_2014, Ghara_2015b, Mondal_2017}. The impact of LC effect is more pronounced at smaller $k$ modes since the cosmic evolution becomes important on the large length-scales along LoS. We also plot the percentage deviations in 3D power spectrum due to LC effect, \textit{i.e.} $\delta_{\rm LC}(\%) = (\Delta^2_{\rm b, CC}/\Delta^2_{\rm b, LC} - 1)\times 100 \%$.

Considering model M$1$ (blue lines) in Fig. \ref{fig:3dps} where X-ray heating is saturated and the signal is driven mainly by the photo-ionization, we note that the deviation $\delta_{\rm LC}$ lies mostly within $\pm 10\%$, except between $z=9.5-12.5$ for $k=0.148~{\rm Mpc}^{-1}$ where $\delta_{\rm LC}$ is as large as $-60\%$ at $z\approx 11$. Our results for M$1$ are qualitatively similar to the earlier LC studies of EoR \citep[e.g.][]{Light_cone_I,La_Plante_2014,Light_Cone_II, Mondal_2017} which demonstrate that the LC effects are important but not as dramatic as we find it. The LC effect is volume-dependent and becomes more(less) pronounced for larger(smaller) volumes. Our choice of such a large volume ($\approx 40~{\rm MHz}$ along LoS) is motivated by the future large-bandwidth surveys by HERA and SKA-Low. The impact of LC effects and $\delta_{\rm LC}$ for M$3$-$f_{\rm X}0.1$ (green lines) are also similar to M$1$ as the X-ray heating here is inefficient ($f_{\rm X}=0.1$). For the more efficient heating ($f_{\rm X} =1$) scenario M$3$-$f_{\rm X}1$, we find that LC effect somewhat enhances $\delta_{\rm LC}$. We find that for M$3$-$f_{\rm X}1$ the clumps of cold \HI~still remain in the IGM (see Fig. \ref{fig:tb_slice_scenario}) where the 21-cm signal is being controlled by the X-ray heating, this allows the deviations to remain important up to $z=7$. The dramatic effect of LC anisotropy on $\Delta^2_{\rm b}(z)$ due to X-ray heating is clearly seen for M$2$-$f_{\rm X}10$ (orange lines) and M$3$-$f_{\rm X}10$ (black lines) where $f_{\rm X} = 10$. We note that the distinct X-ray peak in coeval $\Delta^2_{\rm b}(z)$ (dashed lines) is completely flattened due to LC smoothing. Additionally, a peak appears in the coeval $\Delta^2_{\rm b}(z)$ during the initial stages of the CD ($z \gtrsim 18$) due to fluctuations arising from unsaturated Ly-$\alpha$ coupling. This peak can also get smoothed out in the LC 3D power spectrum. However we could not compute the LC 3D power spectrum due to the limited redshift range of the available LC cube, and thus the peak is not shown in Fig. \ref{fig:3dps}. The deviations here are varying roughly within a range of $-95\%$ to $+88~\%$ and the large deviations are limited to CD and very initial phase of EoR. The LC effects dies down as soon as IGM temperature gets saturated over $\TCMB$ ($z<8$). Note that, our calculations of the 3D PS also includes the pure LoS $k$ modes as well, i.e. when $\kk_\perp =0$, which is not the case in real observations where $\kk_\perp =0$ modes are absent. Therefore, the deviations reported above represents a ceiling for the $\delta_{\rm LC}(\%)$ expected from the observations. However, removing $\kk_\perp=0$ from our analysis does not change the final conclusions and they qualitatively agree with the previous study by \citet{Ghara_2015b} which concludes that the LC effects can not be ignored for the 21-cm signal during CD.

%%%%%%%%%%%%%%%%%%%%%%%%%%%%%%%%%%%%%%%%%%%%%%%%%%%%%%%%%%%%%%%%%%%%%%%%%%
\subsection{Multi-frequency Angular Power Spectrum}\label{sec:estim}
This work aims to quantify the CD-EoR 21-cm signal using power spectrum statistics while considering the inevitable LC effect that makes the signal non-ergodic and aperiodic along the LoS direction. The 3D Fourier modes $\kk$, therefore, will be an inappropriate choice of basis and the widely used 3D power spectrum $\Delta^2_{\rm b}(\kk,z)$ will provide biased estimates of the signal $\dtb(\mathbfit{n}, \nu)$. This restricts utility of the 3D power spectrum estimator to analyzing data in chunks of small bandwidths, thereby limiting access to the large-scale modes along the LoS axis. Moreover, the inference pipelines which compares the 3D power spectra computed from the observed data and coeval 21-cm signal simulations are expected to bias the resultant parametes. In order to claim a fair comparison between simulations and observations, one can properly incorporate the LC effect in inference pipelines. Intuitively one can claim that this approach will avoid the bias in the parameters. However, including the LC effects in the modelling does not rectify the fact that the 3D power spctrum assumes ergodicity along the LoS axis. Thus, the 3D power spectrum is still the incorrect basis for parameter exploration and can still bias the inferred parameter values. One needs to check these claims for the CD-EoR 21-cm signal. The multi-frequency angular power spectrum (MAPS) is a non-stationary two-point statistics which is a suitable alternative for the 3D power spectrum. MAPS, which by definition relaxes the assumption of statistical homogeneity and periodicity along the LoS axis, can completely quantify the two-point statistics of the CD-EoR 21-cm signal observed across a large frequency bandwidth. In general, MAPS is defined by correlating the amplitudes of the orthonormal spherical harmonics $Y_{\ell}^{\rm m}(\mathbfit{n})$ at different observing frequencies (redshifts), after decomposing the signal in a spherical basis set \citep[see eq. 4 of][]{Mondal_2020a} defined on the spherical sky-surfaces corresponding to these frequencies. This decomposition only assumes the signal to be statistically homogeneous and periodic on the spherical sky-surface.

This study further considers the signal to be coming from a small patch in the sky which allows us to work in flat-sky regime where the CD-EoR 21-cm signal can be represented by $\dtb(\tvec, \nu)$ with $\tvec$ being 2D vector on the flat-sky plane. Here we use the 2D Fourier transform $\ttb(\U, \nu)$ of the signal $\dtb(\tvec, \nu)$, where $\U$ (a.k.a. baseline vector) is the Fourier conjugate of $\tvec$. We can also write $\U=\kk_{\perp}/r_{\nu}$ under the flat-sky assumption, where $\kk_{\perp}$ is the component of $\kk$ perpendicular to the LoS direction and $r_{\nu}$ is the comoving distance to the observing frequency. $\ttb(\U, \nu)$ is the primary observed entity of the radio-interferometric observations which can be directly used to compute MAPS without any further processing of the observed signal. Using the flat sky approximation, we can define MAPS as 
\begin{equation}
    \cl(\nu_1, \nu_2) = \Omega^{-1} \,\langle \ttb(\U, \nu_1) \ttb (-\U, \nu_2) \rangle~,
    \label{eq:cl}
\end{equation}
where $\ell = 2\pi |\U|$ and $\Omega$ is the solid angle subtended by the observed (or simulated) sky-patch to the observer. Note that the statistical ergodicity on the sky-patch allows the signal to correlate only at the same baselines $\U$. MAPS does not loose any two-point information if the CD-EoR signal is statistically isotropic on the sky plane. If the signal $\dtb(\tvec, \nu)$ were statistically homogeneous and periodic along the LoS axis, then eq. (\ref{eq:cl}) would have been a function of distance separation along the LoS axis instead of $\nu_1$ and $\nu_2$ separately.

Considering a LC box with typically $\sim 300\times 300 \times 900$ voxels, it will be computationally intractable to store MAPS estimates at each and every combination of cells. Also individual estimates will be too noisy due to statistical fluctuations. In order to circumvent these issues, we compute binned-averaged MAPS by averaging over estimates within a circular bin on $\ell$ plane. We define the binned MAPS estimator for any $i$-th bin as,
\begin{equation}
\begin{split}
    \hat{\mathcal{C}}_{\ell_{i}}(\nu_1, \nu_2) = \frac{1}{2\Omega} \sum_{\U_{{\rm g}_{i}}} w_{{\rm g}_{i}} [& \ttb(\U_{{\rm g}_{i}}, \nu_1) \ttb (-\U_{{\rm g}_{i}}, \nu_2)\\
        &+ \ttb(\U_{{\rm g}_{i}}, \nu_2) \ttb (-\U_{{\rm g}_{i}}, \nu_1)]~,
\end{split}
    \label{eq:bin-cl}
\end{equation}
where the summation is over all the cells $\U_{{\rm g}_{i}}$ in the $i$-th bin and $w_{{\rm g}_{i}} \equiv w(\U_{{\rm g}_{i}})$ denotes the weight corresponding to each cell. The two terms on the right side of eq. (\ref{eq:bin-cl}) and the factor $1/2$ appear because of the symmetry ${\mathcal{C}}_{\ell}(\nu_1, \nu_2) = {\mathcal{C}}_{\ell}(\nu_2, \nu_1)$. Also note that the weight $w_{{\rm g}_{i}}$, in general, can be a function of the two frequencies $(\nu_1,\nu_2)$ as well. In order to compute the bin-averaged MAPS $\bar{\mathcal{C}}_{\ell}(\nu_1,\nu_2)$, we should perform ensemble average of eq. (\ref{eq:bin-cl}), \textit{i.e.} $\bar{\mathcal{C}}_{\ell_{i}}(\nu_1,\nu_2) \equiv \langle \hat{\mathcal{C}}_{\ell_{i}}(\nu_1,\nu_2) \rangle$. However, in this work, we only have single realization of the 21-cm signal which we use to compute $\bar{\mathcal{C}}_{\ell_{i}}(\nu_1,\nu_2)$.

In this work, we use a modified version of the publicly available MAPS code\footnote{\url{https://github.com/rajeshmondal18/MAPS}} which has been used in previous works \citep{Rajesh_Light-cone, Mondal_2019a, Mondal_2020a}. We now motivate and describe the modifications implemented in the existing code. The LC boxes, mentioned in section \ref{sec:sim}, are created by stacking the slices of similar dimensions (in the transverse plane) from the coeval boxes. The LC boxes, thus created, subtend different solid angles at the observer in different channels for being located at a different LoS distances. The nearest slice subtends the largest solid angle and vice-versa. Similarly, the size of each cell on different slices will be scaled accordingly as the number of cells per slice is fixed. This, in turn, produces baseline planes whose extent $U_{\rm max}$ and the cell size $\Delta U$ will vary along the frequency axis. The current available version of code ignores the variation in LoS comoving distance across the frequency band of the LC boxes. It considers only one comoving distance, corresponding to the central frequency of the LC box, while performing the 2D Fourier transform to obtain $\ttb(\U_{\rm g},\nu)$ from $\delta T_{\rm b}(\tvec,\nu)$. Afterwards, it correlates $\ttb(\U_{\rm g},\nu)$ at the same cells in different frequency channels to compute $\cl(\nu_1,\nu_2)$. This approximation is fair only when the frequency difference $|\nu_1-\nu_2|$ is sufficiently small \citep[e.g.][]{Mondal_2019a, Mondal_2020a}. In general, the value $\U_{{\rm g}_{i}}$ corresponding to any particular $i$-th cell will be different for two different frequency channels and therefore the correlation will not provide the estimate of MAPS.

We have modified the existing version of code to resolve this issue by properly treating the variation in LoS distance across the frequency bandwidth of our simulations. The modifications can be listed under the following points.

\begin{itemize}
    \item We first perform 2D Fourier transforms for every slice to obtain $\ttb(\U_{\rm g}, \nu)$ at the respective frequency channel.
    \item We next define a common gridded $uv$ plane whose extent and the cell size remains uniform across different frequency channels. In this case, the extent $U_{\rm max}$ is set by the maximum baseline cell of the nearest slice (smallest redshift) of the LC box as it is the least. The maximum values of the baseline cells increases with the LoS distance. We also fix the cell size of this common $uv$ plane to be the same as that of the farthest frequency slice which is the largest cell size among all frequency channels.
    \item We now interpolate $\ttb(\U_{\rm g}, \nu)$ from the frequency varying $uv$ planes to the common $uv$ plane following the nearest cell method.
    \item Finally we correlate the interpolated $\ttb(\U_{\rm g},\nu)$ on the cells of the common $uv$ planes across the frequency channels to estimate $\cl(\nu_1,\nu_2)$.
\end{itemize}

Note that all the results shown in the following sections have been estimated following the modifications listed above.

%============================================================================================================%

\section{Astrophysical implications}\label{sec:astro}

We compute $\bar{\mathcal{C}}_{\ell_{i}}(\nu_1, \nu_2)$ for the LC boxes simulated using methodology described in section \ref{sec:estim}. Here, we consider the accessible $\boldsymbol{\ell}$-plane by dividing it into $10$ log-spaced circular bins, for which we compute the bin-averaged MAPS (eq. \ref{eq:bin-cl}). Note that we do not perform additional binning of our MAPS estimator along the frequency axis in this paper to retain the evolution information of the signal. However, one can always do channel averaging to reduce the noise in the estimates at the cost of smearing out small-scale information along the LoS axis.

%-------------------------------------------------------------------------------------%
\subsection{Impact of astrophysical processes}

\begin{figure*}
\begin{center}
\includegraphics[scale=0.4]{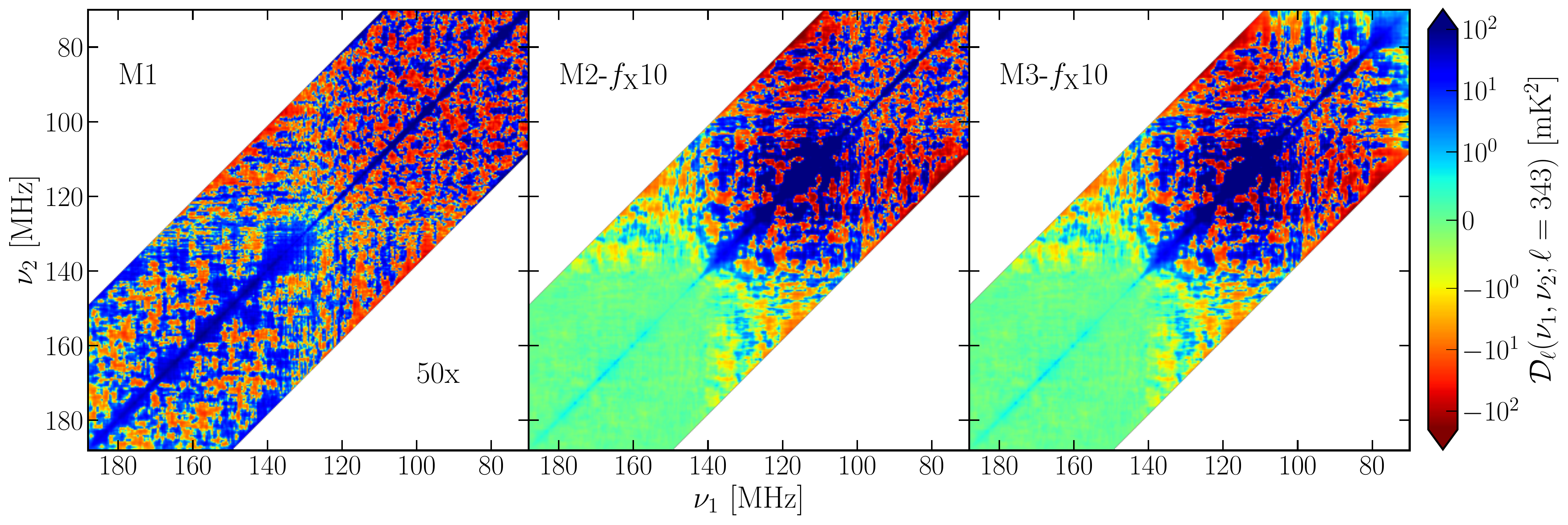}
\includegraphics[scale=0.4]{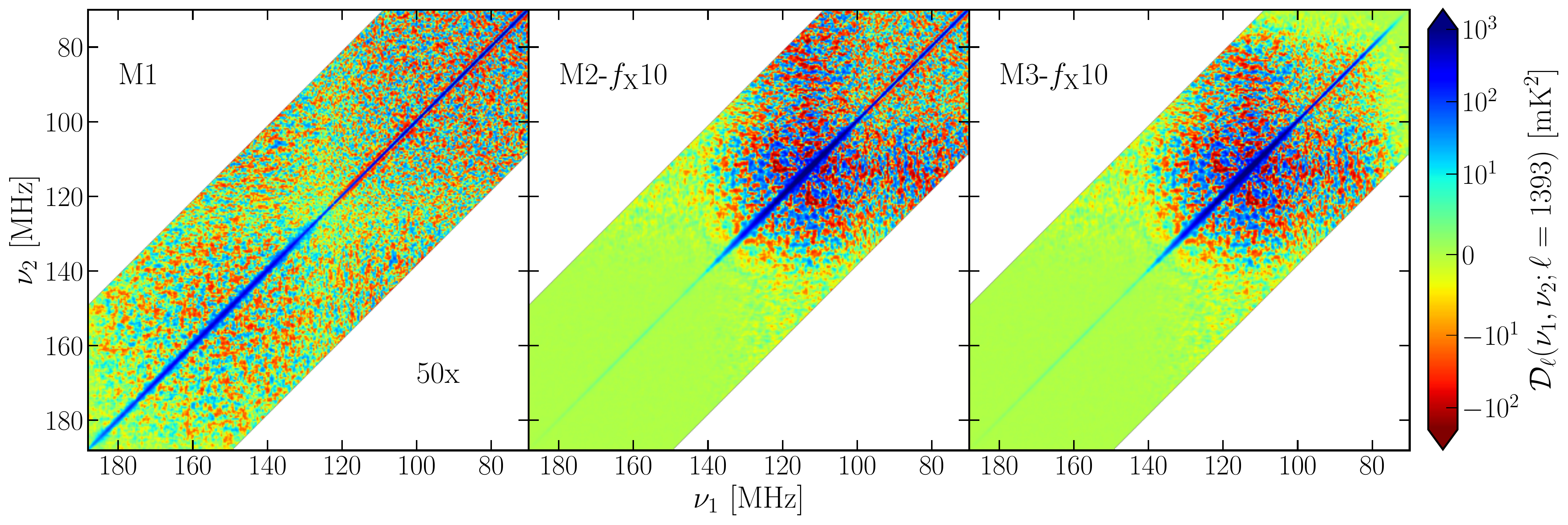}
    \caption{The scaled MAPS $\mathcal{D}_{\ell}(\nu_1,\nu_2)=\ell (\ell+1) \mathcal{C}_\ell(\nu_1,\nu_2)/(2\pi)$ for different CD-EoR models. The three columns here correspond to models M$1$, M$2$-$f_{\rm X}10$ and M$3$-$f_{\rm X}10$, whereas the two rows correspond to the different $\ell$ bins as labelled in the colorbar. We scale the MAPS values for M$1$ by a factor of $50$ to make the patterns visible in the same colorbar.}
   \label{fig:MAPS_full_models}
\end{center}
\end{figure*}

Fig. \ref{fig:MAPS_full_models} shows the scaled bin-averaged MAPS $\mathcal{D}_{\ell}(\nu_1, \nu_2) = \ell(\ell+1)\bar{\mathcal{C}}_{\ell}(\nu_1,\nu_2)/(2\pi)$ for the three different models (in different columns) considered here (see Fig. \ref{fig:tb_slice_model}). The two rows in this figure corresponds to the two $\ell$ bins having mean values $\ell = 343$ and $1393$ which corresponds to respectively large ($63^\prime$) and intermediate ($15.5^{\prime}$) angular-scales on the sky. Here, we show only a part of the MAPS on $(\nu_1,\nu_2)$ plane which are within $300$ channel grid difference that roughly corresponds to $37~{\rm MHz}$ band in frequency. This has to do with the creation of the LC boxes where we simulate signal over $\sim 900$ frequency-grids by repeating slices from the coeval boxes having $[300^3]$ voxels.

\begin{figure}
\begin{center}
\includegraphics[scale=0.31]{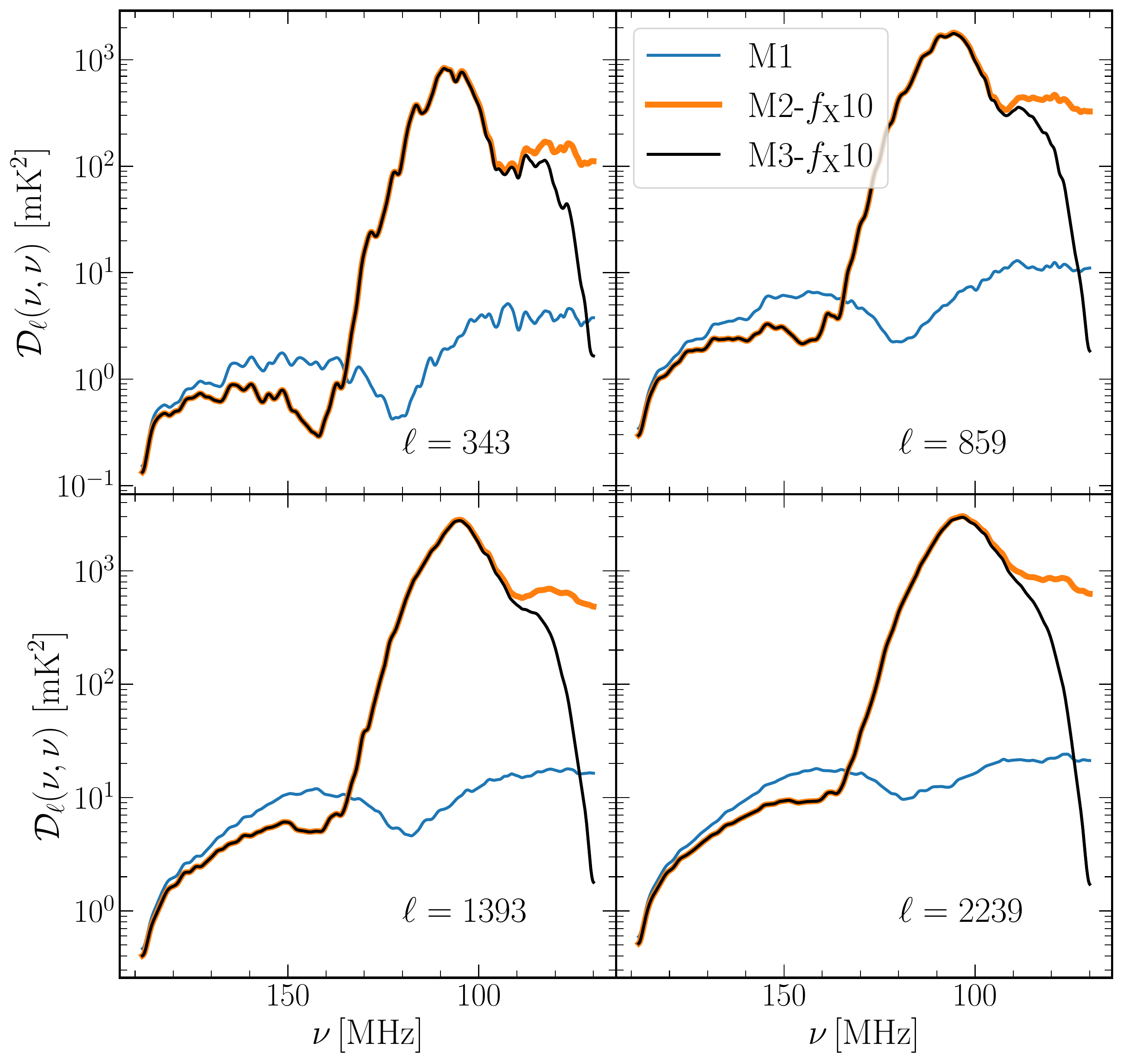}
    \caption{Diagonal of the scaled MAPS $\mathcal{D}_{\ell}(\nu,\nu)=\ell (\ell+1) \mathcal{C}_\ell(\nu,\nu)/(2\pi)$ as a function of $\nu$. The four panels corresponds to the four different $\ell$ bins as labelled. The three colors in each panel refer to the three models M$1$, M$2$-$f_{\rm X}10$ and M$3$-$f_{\rm X}10$ as shown in the legend.}
   \label{fig:MAPS_diag_models}
\end{center}
\end{figure}

We first note that $\mathcal{D}_{\ell}(\nu_1, \nu_2)$ can have both positive and negative values as the signal from different parts in the observation volume can correlate and anti-correlate. The off-diagonal terms contains the information of the LC evolution of the signal as they are the correlations between different the frequency channels. However, the diagonal MAPS $\mathcal{D}(\nu,\nu)$ is always positive and equivalent to 3DPS estimated on $k_{\parallel}=0$ plane. Considering M$1$ where the IGM is heated well above CMB temperature and $T_{\rm s} >> \TCMB$, we find from eq. (\ref{eq:brightnessT}) that the 21-cm signal $\dtb$ is governed by the photo-ionization $\XHI$ during EoR and by density fluctuations $\delta_{\rm B}$ during CD. This, in turn, makes the amplitude of contrast in $\dtb$ (see Fig. \ref{fig:tb_slice_model}), and hence the MAPS, small as seen in the panels of Fig. \ref{fig:MAPS_full_models}. The MAPS estimates in M$1$ are roughly $50-100$ times smaller than the other two models. In contrast with M$1$, we note that the spin-temperature fluctuations due to unsaturated X-ray heating introduces larger fluctuations in $\dtb$ for M$2$-$f_{\rm X}10$ and M$3$-$f_{\rm X}10$ (see Fig. \ref{fig:tb_slice_model}). This boost the MAPS for M$2$-$f_{\rm X}10$ roughly $50-100$ times over that in model M$1$ during the CD. The X-ray heating also makes $\mathcal{D}_{\ell}(\nu_1, \nu_2)$ positive in a larger region around diagonal $(\nu_1=\nu_2)$ during the heating epoch, \textit{i.e.} between $100-130\,{\rm MHz}$. This happens because of the heating bubbles which have been grown to a large enough size before the reionization starts. Before the heating $(\nu < 100 \, {\rm MHz})$, MAPS for M$1$ and M$2$-$f_{\rm X}10$ have similar pattern, however the values are an order of magnitude different. Considering model M$3$-$f_{\rm X}10$, we find the 21-cm MAPS to be the same as that for M$2$-$f_{\rm X}10$ except for the early stages of CD ($\nu < 80~{\rm MHz}$). This is because the unsaturated Ly-$\alpha$ coupling in M$3$-$f_{\rm X}10$ fails to completely couple $T_{\rm s}$ to the gas temperature and $\dtb$ starts near to the zero value during the initial stages of CD. This makes the 21-cm signal and its MAPS weaker for M$3$-$f_{\rm X}10$ than in M$2$-$f_{\rm X}10$. The amplitude of the 21-cm signal and thus its MAPS starts decreasing as soon as the X-ray heating puts $\TS >> \TCMB$ at $\nu \gtrsim 140\,{\rm MHz}$. The signal overall remains positive during EoR (as shown by bluish-green color) on large-scales (small $\ell$) for all the three models considered here. We note that saturated heating of the IGM decreases the 21-cm MAPS by $\sim 2$ orders of magnitudes (see Figs. \ref{fig:3dps} and \ref{fig:MAPS_diag_models}) as compared to the CD for both M$2$-$f_{\rm X}10$ and M$3$-$f_{\rm X}10$.

\begin{figure}
\begin{center}
\includegraphics[scale=0.31]{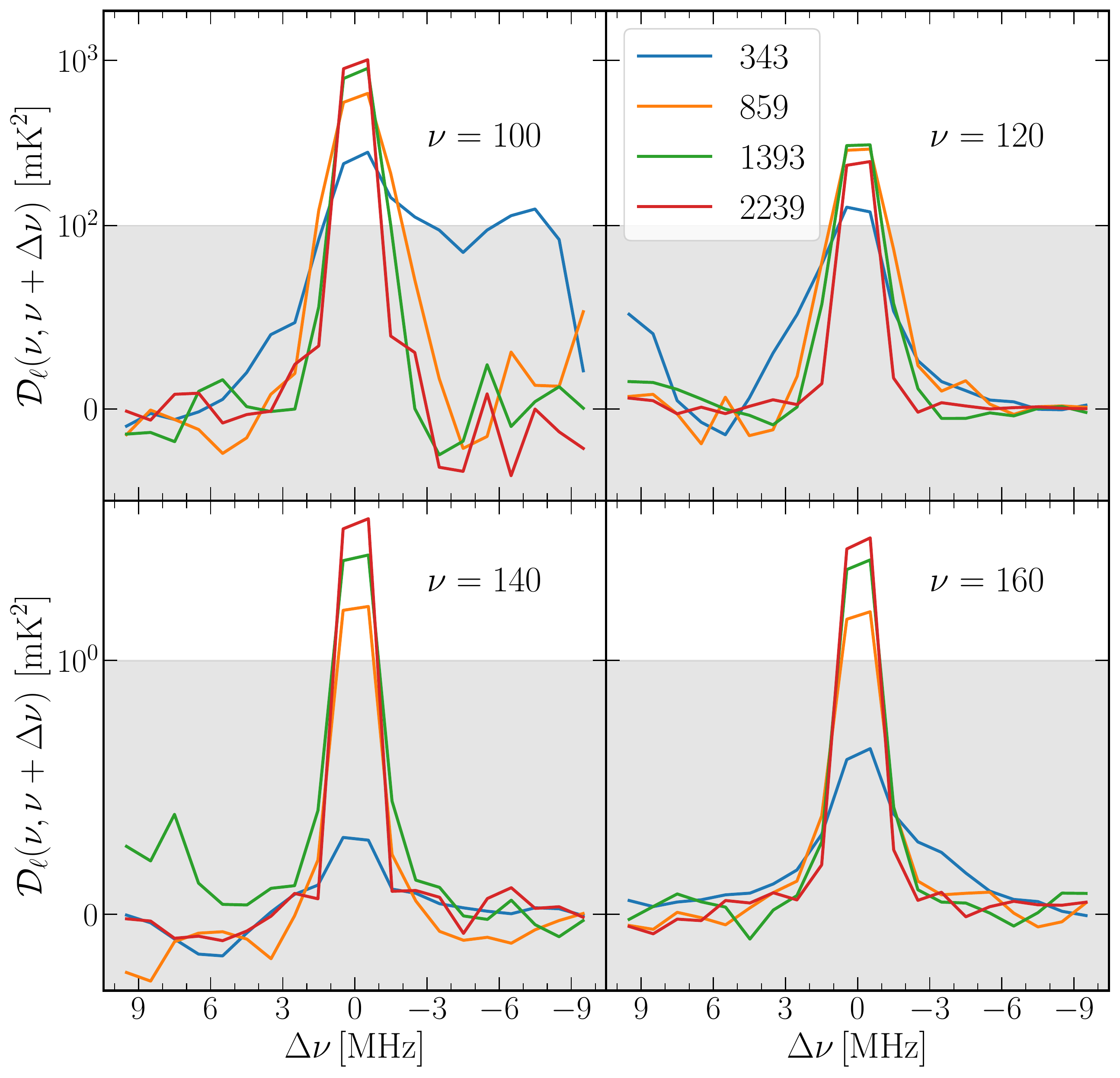}
    \caption{The scaled MAPS $\mathcal{D}_{\ell}(\nu,\nu+\Delta \nu)$ showing correlations as a function of $\Delta \nu$ for M$3$-$f_{\rm X}=10$. The four different panels corresponds to the frequency channels centered at $\nu=100,~120,~140$ and $160~{\rm MHz}$. The four different colors in each panel correspond to the four $\ell$ bins as mentioned in the legend. The gray shades demarcate the linear regime on the symmetric log-scale along vertical axis.}
   \label{fig:MAPS_corr}
\end{center}
\end{figure}

Comparing the top ($\ell = 343$) and the bottom ($\ell = 1393$) rows of Fig. \ref{fig:MAPS_full_models}, we note that the pattern corresponding to each model are roughly identical on the large and intermediate angular-scales respectively. We find that for $\ell =343$, the regions of positive/negative $\mathcal{D}_{\ell}(\nu_1,\nu_2)$ are extended to larger patches on the frequency plane suggesting that the (anti-)correlation between the signal on large-scales does not change rapidly over time. Whereas on small-scales the time-evolution is expected to be faster thereby making the features in $\mathcal{D}_{\ell}(\nu_1,\nu_2)$ smaller for $\ell = 1393$ on frequency plane. We expect the typical sizes of these contiguous patterns to depend on the different stages of CD-EoR and the physical processes. 

\begin{figure}
\begin{center}
\includegraphics[scale=0.32]{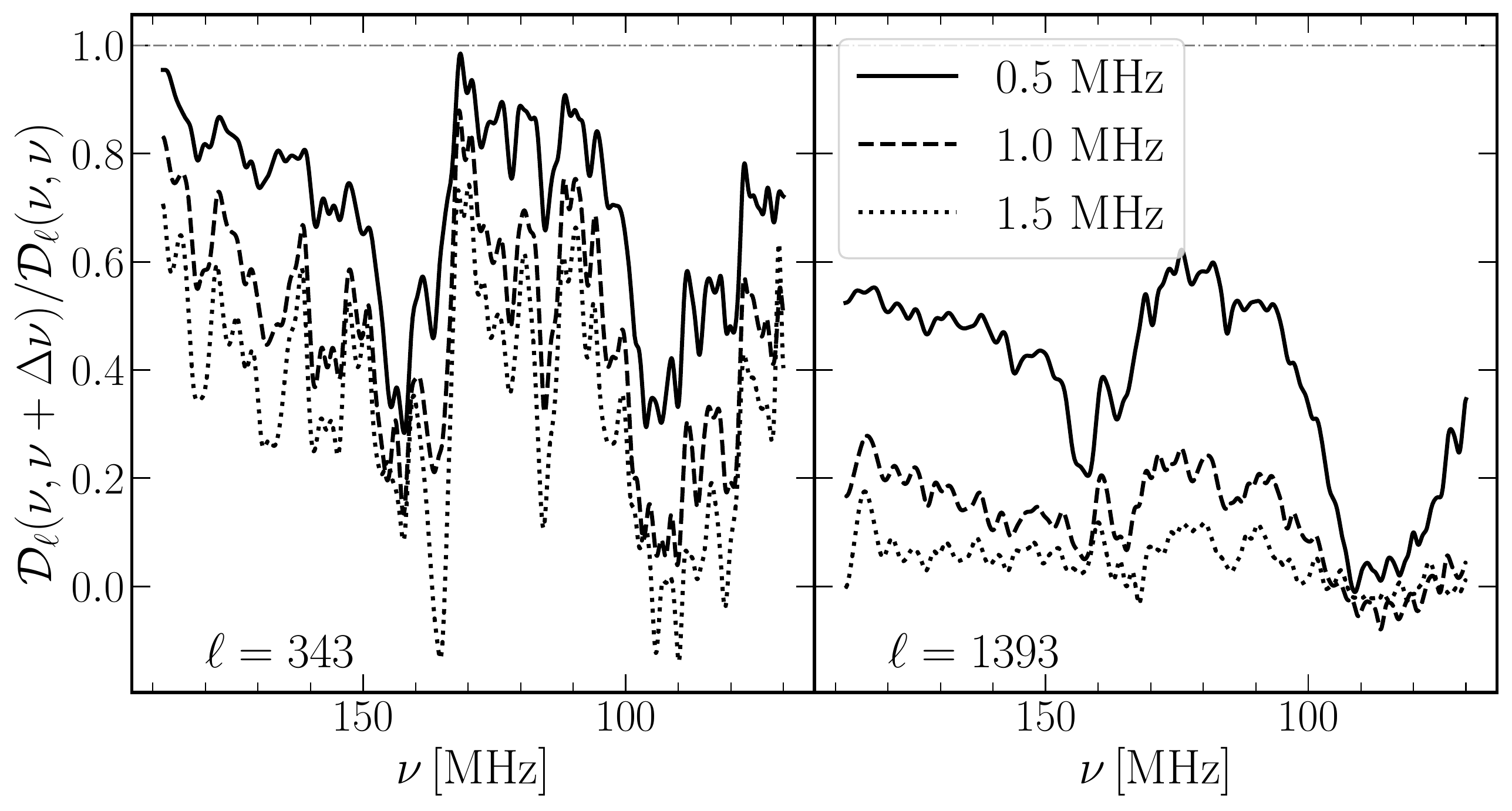}
    \caption{The ratio $\mathcal{D}_{\ell}(\nu,\nu+\Delta \nu)/ \mathcal{D}_{\ell}(\nu,\nu)$ demonstrating how the correlations between MAPS fall when going off the diagonal as a function of $\nu$ for M$3$-$f_{\rm X}=10$. The two different panels corresponds to two different $\ell$ values representing large and intermediate angular-scales on the sky. The three different line styles in each panel correspond to three frequency separations $\Delta \nu = 0.5,~1.0$ and $1.5~{\rm MHz}$. The dot-dashed gray line at the unity represents the case when there is complete correlations.}
   \label{fig:MAPS_offdiag_ratio}
\end{center}
\end{figure} 

Similar to the earlier studies \citep[e.g.,][]{Mondal_2020a}, we find that the 21-cm MAPS takes larger values along and around the diagonal on the $(\nu_1,\nu_2)$ plane. $\mathcal{D}(\nu_1,\nu_2)$ decreases rapidly to one order of magnitude or even less, as we move away from diagonal. This is apparent from Fig. \ref{fig:MAPS_corr} which shows $\mathcal{D}(\nu,\nu+\Delta \nu)$ for our fiducial model (M$3$-$f_{\rm X}10$) as a function of $\Delta \nu$ for four different $\nu=100,~120,~140$ and $160~{\rm MHz}$ corresponding to different stages of CD-EoR in four panels. Different colored lines in each panel corresponds to different $\ell$ values. The rate of decrease at any particular scale ($\ell$) depends upon how fast the signal decorrelates with the frequency separation $|\Delta \nu|$ which in turn depends upon the progression-speed of the underlying physical processes governing the signal in the IGM. We note that the rate at which the signal decorrelates with increasing $|\Delta \nu|$ is slower for smaller $\ell$ values and vice-versa. We expect the behaviour of $\mathcal{D}(\nu, \nu+\Delta \nu)$ to be asymmetrical around $\Delta \nu=0$ as the astrophysical processes may not behave symmetrically around any reference frequency $\nu$. This asymmetry can be more prominent for large-angular scales as can be seen for $\ell=343$ here. The level of asymmetry in the curves will again be dependent on how fast the physical processes evolves the signal at the concerned instant of time. In Fig. \ref{fig:MAPS_offdiag_ratio}, we show the ratio $\mathcal{D}(\nu,\nu+\Delta \nu)/\mathcal{D}(\nu,\nu)$ as a function of $\nu$ for three different values of $\Delta \nu=0.5,~1.0$ and $1.5~{\rm MHz}$ using solid, dashed and dotted lines, respectively. Here, we can clearly see the signal decorrelates at a slower rate wherever X-ray heating determines the 21-cm signal (\textit{i.e.} $\nu \approx 100-140~{\rm MHz}$) for both the $\ell$ values. The decorrelation of the signal is faster during the beginning of CD and the EoR. Therefore, we conclude that a large amount of information of the signal is encoded in the diagonal element of $\mathcal{D}(\nu_1,\nu_2)$. However, in principle, one can also exploit the off-diagonal terms (at least near around the diagonal) while interpreting the signal especially when the X-ray heating is crucial.   

Fig. \ref{fig:MAPS_diag_models} shows the diagonal term $\mathcal{D}_{\ell}(\nu,\nu)$ against four bins centered at $\ell = 343,~859,~1393$ and $2239$ for the three models M$1$, M$2$-$f_{\rm X}10$ and M$3$-$f_{\rm X}10$. We find that the qualitative behaviour is similar to the coeval 3DPS (Fig. \ref{fig:3dps}). This clearly indicates, while analyzing a large-bandwidth signal, MAPS shows the evolutionary features more distinctly than the spherically-averaged 3DPS. However, this comparison is not so straightforward and fair. The advantage of MAPS over 3DPS completely depends on the way we bin and analyze the data. For example, computing the 3DPS by dividing the data into chunks of smaller bandwidths can perform better than when analyzing it in a larger bandwidth. Although, it does not change the fact that there will be LC bias in the estimates due to LoS Fourier transformation. As expected, $\mathcal{D}_{\ell}(\nu,\nu)$ is smaller for M$1$ due to aforementioned reasons. We note that the 21-cm MAPS first decreases as the redshift decreases (or frequency increases). This shallow decrease is because the ionizing radiation wipes out \HI~residing only in the high-density peaks (small-scales) during the initial stages. However, as the reionization progresses further (towards higher frequencies), the fluctuations due to ionized and neutral regions starts increasing the 21-cm MAPS after the first dip at $\nu=120\,{\rm MHz}$ as seen in the figure. $\mathcal{D}_{\ell}(\nu, \nu)$ reaches its maximum when the Universe is $50~\%$ ionized around $\nu=140-150\,{\rm MHz}$. The 21-cm MAPS drops again towards the larger frequencies as the reionization proceeds to the end. Considering models M$2$-$f_{\rm X}10$ and M$3$-$f_{\rm X}10$, the unsaturated heating during CD $(\nu \lesssim 130~{\rm MHz})$ introduces large fluctuations in spin-temperature $T_{\rm s}$ and thereby in the signal. This causes the 21-cm MAPS to increase upto two orders of magnitude in CD regime, when compared with M$1$. The 21-cm MAPS peaks at $\nu \approx 110\,{\rm MHz}$ at all the angular-scales shown here in the four panels. M$2$-$f_{\rm X}10$ and M$3$-$f_{\rm X}10$ are nearly identical except for the very initial stages, i.e. $\nu < 90 \, {\rm MHz}$. In M$3$-$f_{\rm X}10$, the Ly-$\alpha$ coupling is not saturated from the beginning, and thus $T_{\rm s}$ starts at a temperature near to $T_{\gamma}$ resulting into small amplitude of fluctuations and consequently smaller values of 21-cm MAPS. As the IGM evolves (towards higher frequencies), the Ly-$\alpha$ photons couple $T_{\rm s}$ to matter temperature and therefore the amplitude of the 21-cm signal and its MAPS shows a rise in M$3$-$f_{\rm X}10$. However, in M$2$-$f_{\rm X}10$ the Ly-$\alpha$ coupling is saturated, allowing $T_{\rm s} \approx T_{\rm K}$ from the beginning, and therefore the 21-cm MAPS is dominated by the matter fluctuations and remains almost constant till a point ($\nu \approx 90\,{\rm MHz}$) where the X-ray heating starts playing important role in modulating $T_{\rm s}$.

Note that we have chosen $f_{\rm X}=10$ for simulating the 21-cm signal for M$2$-$f_{\rm X}10$ and M$3$-$f_{\rm X}10$. However, to simulate the signal for M$1$, where the heating of IGM is saturated from the beginning of the CD ($\nu= 70\,{\rm MHz}$), we need to choose a larger value for $f_{\rm X}$. Therefore, the heating in M$2$-$f_{\rm X}10$ and M$3$-$f_{\rm X}10$ are comparatively delayed due to which the dip before EoR is occurring at a later redshift (larger frequency) than M$1$. Late heating is also the reason for the mismatch in the value of $\mathcal{D}_{\ell}(\nu,\nu)$ during the initial and middle stages ($\nu \sim 150\,{\rm MHz}$) of reionization. However towards the end of the reionization the heating of IGM is saturated as expected, which in turn causes the MAPS for all the three models to converge. Note that we treat M$3$-$f_{\rm X}10$ as our fiducial model in the rest of this work as this model is closer to realistic scenario based on current theoretical understanding.

%----------------------------------------------------------------------------------------------------------------------------------------------%
\subsection{Impact of X-ray heating efficiencies}

\begin{figure*}
	\begin{center}
		\includegraphics[scale=0.4]{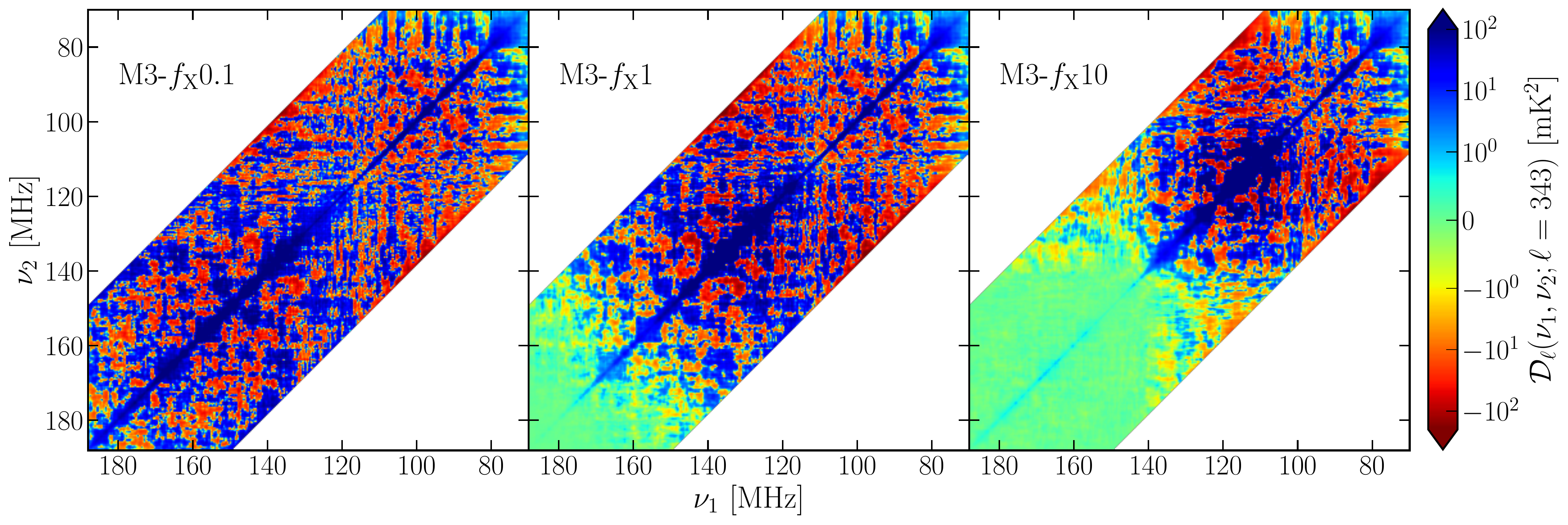}
		\includegraphics[scale=0.4]{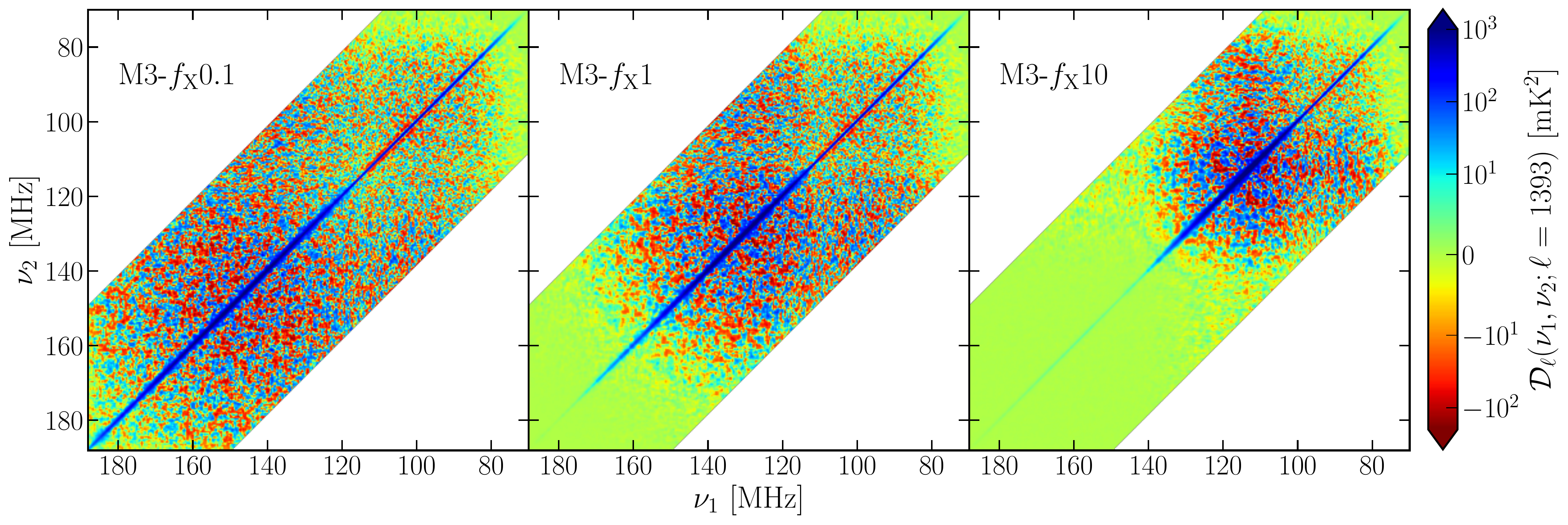}
		\caption{The scaled MAPS $\mathcal{D}_{\ell}(\nu_1,\nu_2)=\ell (\ell+1) \mathcal{C}_\ell(\nu_1,\nu_2)/(2\pi)$ for different heating scenarios. The three columns here are for three heating efficiencies $f_{\rm X}=0.1,\,1$ and $10$ corresponding to the self-consistent model M$3$. The two rows correspond to the different $\ell$ bins as labelled in the colorbar.}
		\label{fig:MAPS_full_scenario}
	\end{center}
\end{figure*}

\begin{figure}
\begin{center}
\includegraphics[scale=0.31]{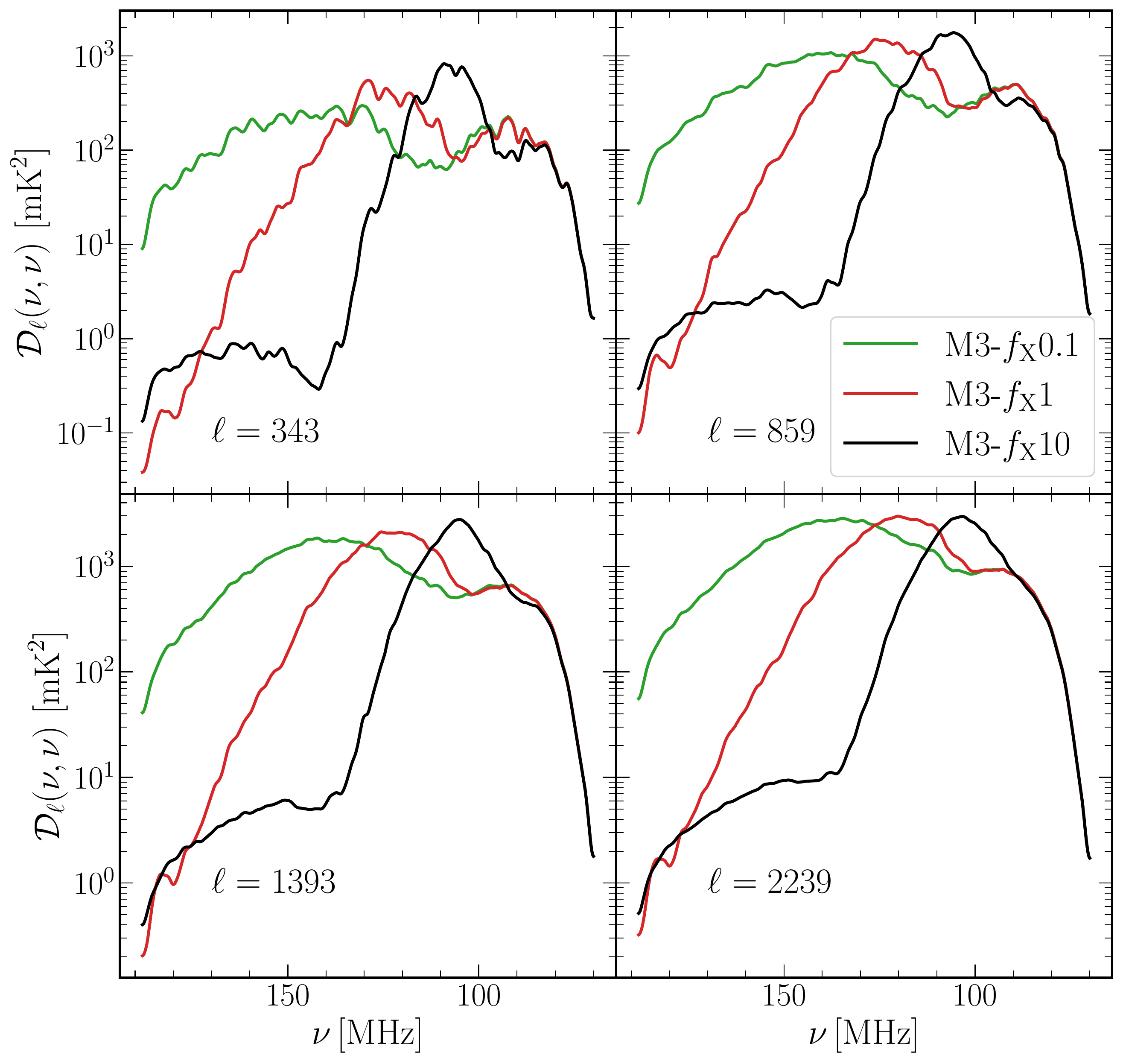}
    \caption{Diagonal of the scaled MAPS $\mathcal{D}_{\ell}(\nu,\nu)=\ell (\ell+1) \mathcal{C}_\ell(\nu,\nu)/(2\pi)$ as a function of $\nu$. The four panels correspond to the four different $\ell$ bins as labelled. The three colors in each panel are for three heating efficiencies $f_{\rm X}=0.1,\,1$ and $10$ corresponding to model M$3$.}
   \label{fig:MAPS_diag_scenario}
\end{center}
\end{figure}

We further study the effects of different heating efficiencies $f_{\rm X} = 0.1,~1$ and $10$ on 21-cm MAPS for our fiducial model~M$3$. The signal for the three scenarios is expected to be similar until X-ray heating starts playing important role in modulating $T_{\rm s}$. This behaviour is apparent from Fig. \ref{fig:MAPS_full_scenario} which shows $\mathcal{D}_{\ell}(\nu_1,\nu_2)$ for M$3$-$f_{\rm X}0.1$, M$3$-$f_{\rm X}1$ and M$3$-$f_{\rm X}10$. We clearly see that pattern on the $(\nu_1,\nu_2)$ plane is qualitatively similar for all the three scenarios as long as the Ly-$\alpha$ coupling plays important role. We also note that the 21-cm MAPS becomes strictly positive over a large region near the diagonal ($\nu_1 \approx \nu_2$) once the X-ray heating or the ionizing photons starts becoming important for the signal. The frequency at which this happens decreases with the increase in $f_{\rm X}$ value. Also, the IGM gets heated to a saturation level quickly for larger $f_{\rm X}$ values and the 21-cm MAPS decreases at a faster rate towards larger frequencies (lower redshifts) as seen in Figs. \ref{fig:MAPS_full_scenario} and \ref{fig:MAPS_diag_scenario}.

We plot the diagonal MAPS $\mathcal{D}_{\ell}(\nu, \nu)$ in Fig. \ref{fig:MAPS_diag_scenario}, where the four panels again corresponds to the four bins centered at $\ell = 343,~859,~1393$ and $2239$. Considering M$3$-$f_{\rm X}0.1$, we find that the heating of the IGM is inefficient here and we still find cold \HI~clouds till the end of the EoR (see top panel of Fig. \ref{fig:tb_slice_scenario}). For this reason the heating peak is absent in this scenario. The second peak (at larger frequency) here is due to the reionization. The behaviour of 21-cm MAPS here is qualitatively similar to that in M$1$ except for its values which are orders of magnitude larger. We find the heating peak of 21-cm MAPS exists for both M$3$-$f_{\rm X}1$ and M$3$-$f_{\rm X}10$. However the peak for M$3$-$f_{\rm X}1$ occurs at a later redshift (larger frequency) than that in M$3$-$f_{\rm X}10$. We also note that the reionization peak in 21-cm MAPS is absent for M$3$-$f_{\rm X}1$ as the IGM was not heated sufficiently for the signal to get governed majorly by the ionization field. This is not the case for M$3$-$f_{\rm X}10$, for which the heating of IGM gets saturated before the reionization starts and we do see a reionization peak in this scenario. The dpendence of the CD-EoR 21-cm MAPS on heating efficiencies is qualitatively similar to what have been observed in the 3D power spectra computed from the coeval boxes of CD-EoR 21-cm signal in previous studies \citep[e.g.][]{Mesinger_2013, Ghara_2015a, Shimabukuro_2015}.

%============================================================================================================%

\section{Detectability}\label{sec:detect}
At first, we derive the formula for the error covariance of the MAPS statistics here. Later, we use the formula to predict the detectability of the diagonal component of the CD-EoR 21-cm MAPS, i.e. $\mathcal{D}_{\ell}(\nu,\nu)$, in the context of observations using HERA, NenuFAR and the upcoming SKA-Low.

\subsection{MAPS error covariance}\label{subsec:cov}

Any radio-interferometer directly measures the CD-EoR 21-cm signal visibilities $\ttb(\U, \nu)$ at frequency, say, $\nu$ and baselines $\U=\mathbfit{d}/\lambda$, where $\mathbfit{d}$ is the antenna-pair separation projected onto the sky-plane, perpendicular to a chosen central LoS, and $\lambda$ is the wavelength corresponding to $\nu$. The baseline tracks in different frequency channels are the points where the incoming signal is being sampled. The system noise, which originates from statistical fluctuations in the sky and the receivers, is inherent to any radio-interferometric measurement. Additionally, the foreground and other systematic errors corrupt the actual cosmological signal.
These contamination affects the MAPS estimate as well as the corresponding error in the measurement. However, in this analysis, we assume that the foreground is perfectly modelled and completely removed from the data and also that systematic errors have been removed. Therefore the measured visibilities will be the sum of cosmological 21-cm signal and the system noise, \textit{i.e.} $\tilde{T}_{\rm t2}(\U, \nu) = \ttb(\U, \nu) + \tilde{T}_{\rm N2}(\U, \nu)$. The system noise is a Gaussian random field on sky-surface and hence $\tilde{T}_{\rm N2}(\U, \nu)$ can be assumed to be uncorrelated at different baselines and frequencies. Therefore according to eq. (\ref{eq:cl}), the MAPS of the total observed signal will then be the sum of signal MAPS $\cl(\nu_1, \nu_2)$ and the system noise MAPS $\cl^{\rm N}(\nu_1, \nu_2)$, \textit{i.e.}, $\cl^{\rm t} (\nu_1, \nu_2)= \cl(\nu_1, \nu_2)+ \delta^{\rm K}_{\nu_1,\nu_2} \cl^{\rm N}(\nu_1, \nu_2)$. The Kroneker's delta $\delta^{\rm K}_{\nu_1,\nu_2}$ appears here due to the fact that the Gaussian system noise at different frequencies remains uncorrelated. Although it is possible to remove the noise bias from the total MAPS estimate \citep{Bharadwaj_2018}, its contribution to the total measurement error still remains.

Ignoring the non-Gaussianity of the underlying 21-cm field, we have calculated the error covariance matrix of the bin-averaged MAPS estimator (eq. \ref{eq:bin-cl}) which, in general, can be written as \citep[see Appendix A of][for the derivation]{Mondal_2020a}
\begin{equation}
    \begin{split}
        \mathbfit{X}_{12,34}^{\ell_{i}} &= \langle \delta \hat{\mathcal{C}}_{\ell_{i}}^{\rm t}(\nu_1,\nu_2) ~\delta \hat{\mathcal{C}}_{\ell_{i}}^{\rm t}(\nu_3,\nu_4)\rangle \\
        &=\frac{1}{2} \sum_{\U_{{\rm g}_{i}}} w_{{\rm g}_{i}}^2 \left[\mathcal{C}_{\ell_{{\rm g}_{i}}}^{\rm t}(\nu_1, \nu_3) \mathcal{C}_{\ell_{{\rm g}_{i}}}^{\rm t}(\nu_2, \nu_4) \right.\\ 
        &\left. \qquad \qquad \qquad + \mathcal{C}_{\ell_{{\rm g}_{i}}}^{\rm t}(\nu_1, \nu_4) \mathcal{C}_{\ell_{{\rm g}_{i}}}^{\rm t}(\nu_2, \nu_3)\right]~,
    \end{split}
    \label{eq:errorcov}
\end{equation}
where $\delta \hat{\mathcal{C}}_{\ell_{i}}^{\rm t}$ denotes the deviation of $\hat{\mathcal{C}}_{\ell_{i}}^{\rm t}$ from its ensemble mean and $w_{{\rm g}_{i}}$ is the weight corresponding to each cell and the summation is over the cells $\U_{{\rm g}_{i}}$ within the $i$-th bin. It is apparent from eq. (\ref{eq:errorcov}) that the errors in different baseline cells $\U_{\rm g}$ are completely uncorrelated here. This is because of our Gaussian assumption for both the signal and the system noise field. However, the errors may be correlated for two different pairs of frequencies, i.e. ($\nu_1, \nu_2$) and ($\nu_3, \nu_4$). In reality, the CD-EoR 21-cm signal is a non-Gaussian field \citep[e.g.][]{Bharadwaj_Pandey, Harker_2009a} due to non-Gaussianity arising due to various factors such as $\XHI,~\delta_{\rm B},~T_{\rm s}$ in the signal (eq. \ref{eq:brightnessT}). This non-Gaussianity gives rise to the non-zero contributions from angular trispectrum to the cosmic variance (CV) of MAPS and also introduces correlations between different baselines \citep[e.g.][]{Mondal_2015, Mondal_I}. \citet{Mondal_I} and \citet{Mondal_II} have developed a novel technique to estimate the 3D trispectrum from a moderately large ensemble of 21-cm simulations. One can extend their technique to estimate the angular trispectrum for the CD-EoR 21-cm signal, although it will be more computationally challenging. Hence, we drop the non-Gaussian contribution in our calculations and analysis. Recently, \citet{shaw_2019} have studied the importance of trispectrum contribution to the total error prediction in 3D power spectrum in the context of future SKA-Low observations. Following their results, we argue that non-Gaussian effects will be important for the small and intermediate $\ell$ bins where the system noise contribution is relatively smaller.

\begin{table*}
	%\centering
	\tabcolsep 6pt
	\renewcommand\arraystretch{1.5}
	\caption{Observational parameters for different radio-interferometers used to compute the system noise MAPS $\cl^{\rm N}(\nu,\nu)$ in this analysis.}
	\begin{tabular}{c c c c c c c c c c}
		\hline
		Telescopes & No. of & Diameter & Frequency Bandwidth & Channel width &$T_{\rm rec}$  & Pointing & $\ell_{\rm max}$ & $\Delta \ell$ & Survey\\ 
		& stations & (${\rm m}$) & (${\rm MHz}$) & (${\rm kHz}$) & (${\rm K}$) & direction & & & mode\\
		\hline
		\hline
		HERA & $350$ & $14$ & $50-250$ & $97.66$ & $170$ & $-30^{\circ}43^{\prime}17^{\prime\prime}$ & $11250$ & $97.82$ & Drift Scan \\
		NenuFAR & $102$ & $25$ & $10-85$ & $195.32$ & $1000$ & $+47^{\circ}22^{\prime}34^{\prime\prime}$ & $12093$ & $108.95$ & Field Track \\
		SKA-Low & $512$ & $35$ & $50-350$ & $97.66$ & $100$ & $-26^{\circ}49^{\prime} 29^{\prime\prime}$ & $11250$ & $97.82$ & Field Track \\
		\hline
	\end{tabular}
\begin{flushleft}
	{$^{\boldsymbol{\ddagger \ddagger}}$ Here $T_{\rm rec}$ denotes the antenna receiver temperature in ${\rm K}$. $\ell_{\rm max}$ denotes the maximum extent of the square $\ell$ plane from its center and $\Delta \ell$ denotes the corresponding cell size, considered for the noise predictions. Note that we considers observations per night to be of $8~{\rm h}$.}
	%\caption{Lists parameters for different radio-interferometers used to compute the system noise MAPS $\cl^{\rm N}(\nu,\nu)$ in this analysis that considers observations of $8~{\rm h/night}$. Here $T_{\rm rec}$ denotes the antenna receiver temperature in ${\rm K}$. $\ell_{\rm max}$ denotes the maximum extent of the square $\ell$ plane from its center and $\Delta \ell$ denotes the corresponding cell size, considered for the noise predictions.}
\end{flushleft}
	\label{tab:params}
\end{table*}

Here we simplify the full error covariance matrix and restrict ourselves to predicting the error variance of our MAPS estimator which can be written as
\begin{equation}
    \begin{split}
        \mathbfit{X}_{12,12}^{\ell_{i}} &= [\boldsymbol{\sigma}_{12}^{\ell_{i}}]^2 = \langle [\delta \hat{\mathcal{C}}_{\ell_{i}}^{\rm t}(\nu_1,\nu_2)]^2 \rangle \\
        &=\frac{1}{2} \sum_{\U_{{\rm g}_{i}}} w_{{\rm g}_{i}}^2 \left[\mathcal{C}_{\ell_{{\rm g}_{i}}}^{\rm t}(\nu_1, \nu_1) \mathcal{C}_{\ell_{{\rm g}_{i}}}^{\rm t}(\nu_2, \nu_2) + \{ \mathcal{C}_{\ell_{{\rm g}_{i}}}^{\rm t}(\nu_1, \nu_2) \}^2 \right]\\
        &=\frac{1}{2}\sum_{\U_{{\rm g}_{i}}} w_{{\rm g}_{i}}^2 \left[\{\mathcal{C}_{\ell_{{\rm g}_{i}}}(\nu_1, \nu_1)+\mathcal{C}_{\ell_{{\rm g}_{i}}}^{\rm N}(\nu_1, \nu_1)\} \right.\\  
        & \qquad \qquad \qquad \times \{\mathcal{C}_{\ell_{{\rm g}_{i}}}(\nu_2, \nu_2)+\mathcal{C}_{\ell_{{\rm g}_{i}}}^{\rm N}(\nu_2, \nu_2)\} \\
        &\left.  \qquad \qquad \qquad + \{\mathcal{C}_{\ell_{{\rm g}_{i}}}(\nu_1, \nu_2)+\delta^{\rm K}_{\nu_1,\nu_2} \mathcal{C}_{\ell_{{\rm g}_{i}}}^{\rm N}(\nu_1, \nu_2)\}^2~\right].
    \end{split}
    \label{eq:var12}
\end{equation}

Hence, we need the weights $w_{\rm g}$, the 21-cm MAPS $\mathcal{C}_{\ell_{\rm g}}(\nu_1,\nu_2)$ and the system noise MAPS $\mathcal{C}_{\ell_{\rm g}}^{\rm N}(\nu_1,\nu_2)$ at every $\boldsymbol{\ell}_{\rm g}$ cell to predict the error variance in 21-cm MAPS. We have estimated the 21-cm bin-averaged MAPS using simulated signal LC boxes. Note that the bin-averaged MAPS presented above in sec. \ref{sec:astro} considers uniform weights across the cells while estimating $\bar{\mathcal{C}}_{\ell_{i}}(\nu_1,\nu_2)$ from simulated signal. 

Next, we compute the system noise MAPS at cells $\U_{\rm g} = \boldsymbol{\ell}_{\rm g}/(2\pi)$ using \citep[e.g.][]{White_1999, ZFH_2004, Mondal_2020a} 
\begin{equation}
    \mathcal{C}_{\ell_{\rm g}}^{\rm N}(\nu,\nu) = \frac{8 \,{\rm h}}{t_{\rm obs} \,\tau_8(\U_{\rm g})} \times \frac{T_{\rm sys}^2 \, \lambda^4}{N_{\rm p} \, \Delta t \, \Delta \nu \, a^2 \int d\U^\prime |\tilde{A}(\U - \U^\prime)|^2}~,
    \label{eq:CN}
\end{equation}
where $\nu$ is the observing frequency and $\Delta \nu$ is the corresponding channel width. In the equation above, $t_{\rm obs}$ is the total observation time and $\tau_8(\U_{\rm g})$ denotes the gridded baseline distribution considering $uv$ tracks of $8$ h/night observations. One need to adjust the $8\,{\rm h}$ in the numerator in case the baseline tracks have been simulated for different observation hours per night. Here we consider the system temperature $T_{\rm sys}$ to be a sum of sky temperature $T_{\rm sky} = 60 \lambda^{2.55}\,{\rm K}$ \citep[e.g.][]{Reich_1988, Platania_1998, Rogers_2008, Bernardi_2009, Fixsen_2011, McKinley_2018, Mozdzen_2018, Spinelli_2021} and the receiver temperature $T_{\rm rec}$ \citep[e.g.][]{shaw_2019, Mondal_2020a}. Note that the $\lambda$ considered in $T_{\rm sky}$ is in meters and $T_{\rm rec}$ is the amplitude of the white Gaussian noise generated by the antenna receiver. $N_{\rm p}=2$ is number of dipole polarisation, $\Delta t = 60\,{\rm sec}$ is the integration time for each baseline, $\Delta \nu$ is the frequency channel width, $a^2$ is the physical area of each antenna element and $\tilde{A}(\U)$ is the Fourier transform of the antenna beam pattern $A(\tvec)$. For simplicity, we consider each antenna element to be circular with a diameter $D$. Considering circular symmetry, we approximate the main lobe of the primary beam by a Gaussian $A(\theta)= \exp(-\theta^2/\theta_0^2)$ \citep{Choudhuri2014, Mondal_2020a}. The computation of system noise MAPS (eq. \ref{eq:CN}) at every $\U_{\rm g}$ cell depends on various design specifications of the telescopes which are tabulated in Table \ref{tab:params}.

We now consider a non-uniform weighting scheme as the system noise contribution varies across the cells $\U_{\rm g}$ for the noisy observational data. We obtain the weights at every cells by maximizing the signal-to-noise ratio which is defined as ${\rm SNR}_{12}^{\ell_{i}} \equiv \bar{\mathcal{C}}_{\ell_{i}}(\nu_1,\nu_2)/\boldsymbol{\sigma}_{12}^{\ell_{i}}$. We make a simplifying assumption that the 21-cm MAPS does not vary significantly across the cells within a particular bin, \textit{i.e.} $\mathcal{C}_{\ell_{{\rm g}_{i}}}(\nu_1,\nu_2) = \bar{\mathcal{C}}_{\ell_{i}}(\nu_1,\nu_2)$. Under this assumption, we find that the unnormalized weights can be written as \citep[see e.g.,][]{Feldman_1994, Mondal_2020a}
\begin{equation}
\begin{split}
    \hat{w}_{{\rm g}_{i}} = [ \{ \bar{\mathcal{C}}_{\ell_{i}}(\nu_1, \nu_1) +& \mathcal{C}^{\rm N}_{\ell_{{\rm g}_{i}}}(\nu_1, \nu_1) \}
    \{ \bar{\mathcal{C}}_{\ell_{i}}(\nu_2, \nu_2) + \mathcal{C}^{\rm N}_{\ell_{{\rm g}_{i}}}(\nu_2, \nu_2) \} \\
    & +\{ \bar{\mathcal{C}}_{\ell_{i}}(\nu_1, \nu_2) + \delta^{\rm K}_{\nu_1,\nu_2} \mathcal{C}^{\rm N}_{\ell_{{\rm g}_{i}}}(\nu_1, \nu_1)\}^2 ]^{-1}~,
\end{split}
\label{eq:wt}
\end{equation}
where $w_{{\rm g}_{i}}=\hat{w}_{{\rm g}_{i}}/\sum_{\U_{{\rm g}_{i}}} \hat{w}_{{\rm g}_{i}}$. The weight (eq. \ref{eq:wt}) here varies inversely to square of system noise power spectrum $\mathcal{C}^{\rm N}_{\ell}(\nu,\nu)$, hence cells with larger noise contaminations will contribute less to the bin-averaged MAPS and vice-versa. The cells which do not have $uv$ sampling (\textit{i.e.} $\tau_8(\U_{\rm g}) =0$) equivalently have infinite noise (eq. \ref{eq:CN}). Using the weight in eq. (\ref{eq:var12}) the variance in $\bar{\mathcal{C}}_{\ell_{i}}(\nu_1, \nu_2)$ can be written as
\begin{equation}
    [\boldsymbol{\sigma}_{12}^{\ell_{i}}]^2 = \frac{1}{2} \times \frac{1}{\sum_{\U_{{\rm g}_{i}}} \hat{w}_{{\rm g}_{i}}}~.
    \label{eq:var12_opt}
\end{equation}
Next, considering the diagonal elements where MAPS $\bar{\mathcal{C}}_{\ell_{i}}(\nu, \nu)$ peaks, the corresponding variance (eq. \ref{eq:var12}) can further be reduced down to 
\begin{equation}
    [\boldsymbol{\sigma}_{11}^{\ell_{i}}]^2 = \frac{1}{\sum_{\U_{{\rm g}_{i}}} [ \bar{\mathcal{C}}_{\ell_{i}} (\nu_1,\nu_1) + \mathcal{C}_{\ell_{{\rm g}_{i}}}^{\rm N} (\nu_1,\nu_1)]^{-2}}~.
    \label{eq:var11_opt}
\end{equation}
We finally use eq. (\ref{eq:var11_opt}) to compute the error variance in the measurements of $\mathcal{D}_{\ell}(\nu,\nu)$ for our fiducial model (M$3$-$f_{\rm X}10$). We present the results for three current and upcoming radio-interferometers HERA, NenuFAR and SKA-Low below.
%----------------------------------------------------------------------------------------------------------------------------------%

\subsection{Detectability with HERA}\label{subsec:hera}
\begin{figure*}
\begin{center}
\includegraphics[scale=0.5]{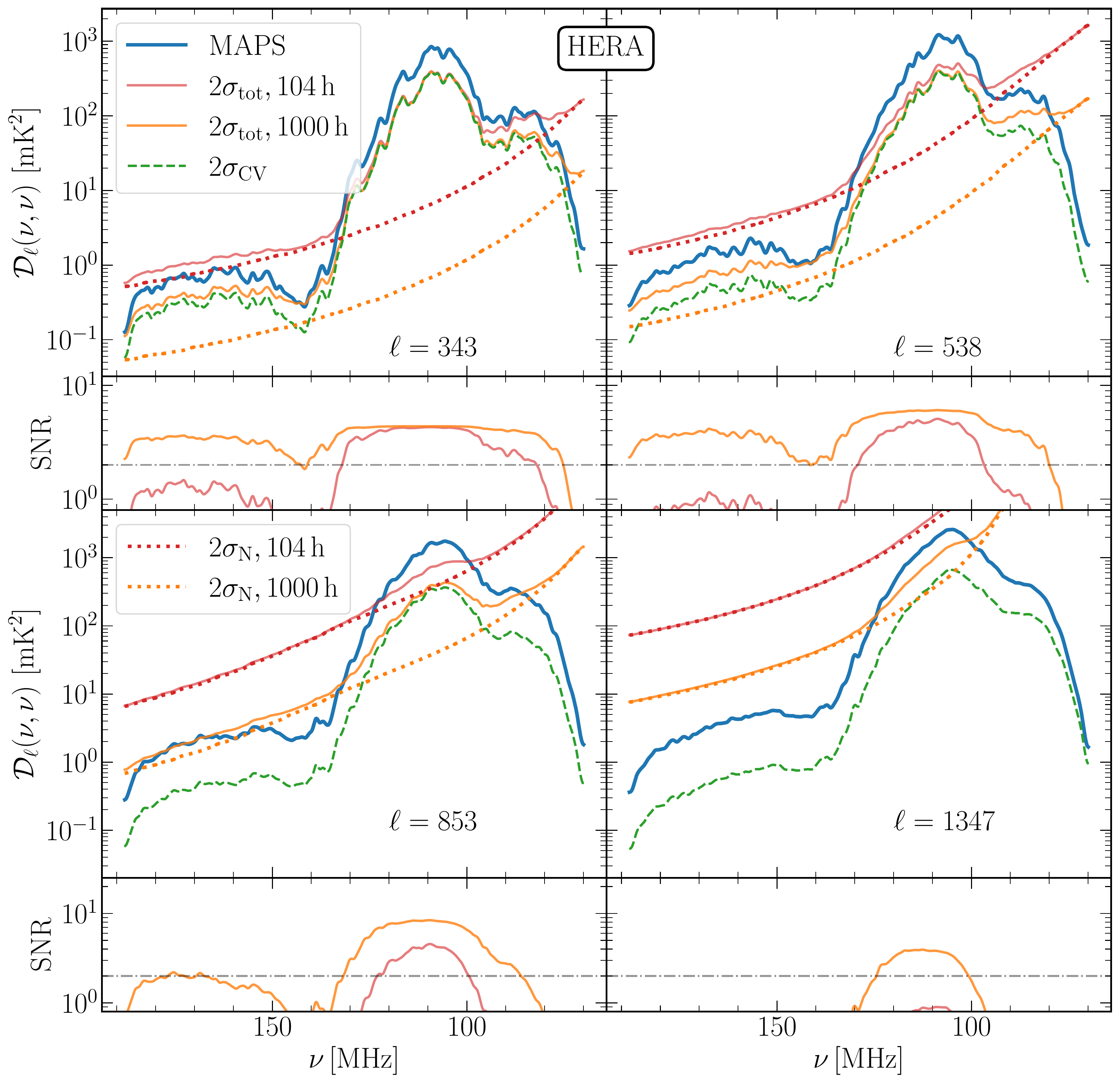}
    \caption{HERA predictions for detecting the diagonal $(\nu_1=\nu_2)$ of the scaled MAPS $\mathcal{D}_{\ell}(\nu,\nu)=\ell (\ell+1) \mathcal{C}_\ell(\nu,\nu)/(2\pi)$. The four panels here corresponds to the different $\ell$ bins as labelled. The thick solid blue lines show the expected CD-EoR 21-cm MAPS as a function of $\nu$ corresponding to M$3$-$f_{\rm X}10$. The red and orange solid lines in the four bigger panels represents the total $2\sigma$ error estimates considering $97.66~{\rm KHz}$ channel-width, respectively for $104\,{\rm h}$ and $1000\,{\rm h}$ of HERA drift-scan observations. Corresponding dotted red and orange lines represents the system noise contribution ($2\sigma$) to the total error for $104\,{\rm h}$ and $1000\,{\rm h}$ respectively. The dashed green lines in the four bigger panels represents the respective CV error estimates ($2\sigma$). The four smaller panels, corresponding to the bottom of each bigger panel, show how SNR of the measurements varies with the observing frequency $\nu$ for $104\,{\rm h}$ (red) and $1000\,{\rm h}$ (orange). The gray dashed lines in the smaller panels demarcates SNR$=2$ floor for detection.}
   \label{fig:HERA_pred}
\end{center}
\end{figure*}

 The Hydrogen Epoch of Reionization Array (HERA\textsuperscript{\ref{ft:hera}}) is a radio-interferometric telescope located in the Karoo desert, South Africa. It consists of zenith pointing dishes of $14~{\rm m}$ diameter which are arranged in a closed packed hexagonal core of about $300~{\rm m}$ across. Currently this telescope is observing in its first phase with $50$ dishes and within frequency bandwidth $100-200~{\rm MHz}$ that spans EoR. In this analysis, we consider the whole frequency bandwidth is divided into $1024$ channels having channel width $97.66~{\rm kHz}$. Recently, \citet{Abdurashidova_2022a, Abdurashidova_2022b} have put best upper limits to date on EoR 21-cm 3D power spectrum by analysing roughly $1000~{\rm h}$ of HERA Phase I data. In its final phase, the array will consists of total $350$ dishes all observing within a larger frequency range $50-250~{\rm MHz}$ which covers both the CD and EoR \citep[see e.g.][]{Fagnoni_2021}. We consider the upcoming final phase of HERA for our predictions here. We simulate baseline tracks of HERA observing in drift scan mode at a declination DEC$= -30^{\circ} 43^{\prime} 17^{\prime \prime}$ (phase center of the telescope). We finally compute the total error variance $[\boldsymbol{\sigma}_{12}^{\ell_{i}}]^2$ using eq. (\ref{eq:var12_opt}) where we set $T_{\rm rec} = 170~{\rm K}$ (see Table \ref{tab:params}) which is similar to that of the Phase I receivers \citep[see e.g.][]{Thyagarajan_2020, Fagnoni_2021}. Note that the common square $\ell$ plane which is used for HERA predictions extends between $\pm 11250$ with cell size being $\Delta \ell = 97.82$. We divide this accessible $\ell$ plane into $10$ log-spaced circular bins for estimating the errors.

Unlike field tracking, the sky patch which HERA is looking at constantly changes with time in drift-scan mode. Thus, we can say that at every integration time-stamp HERA measures a different realization of the signal. This gives rise to additional CV in the 21-cm MAPS measured at every voxel. Based on this intuition, we add the contribution of this excess variance in quadrature to eq. (\ref{eq:var11_opt}) for simplicity. Assuming the fluctuations in the MAPS due to drifting sky patch to be Gaussian, we expect the additional variance in the diagonal MAPS to be proportional to the average measurement, \textit{i.e.} $\mathcal{D}^2_{\ell}(\nu,\nu)/N_{\rm snap}$. The number of snapshots of the different sky patch drifting per night of observations $N_{\rm snap}$ is set by the integration time of the interferometer. Here we consider HERA observing $8$ hours per night with $60~{\rm sec}$ integration time resulting into $N_{\rm snap} =480$. One can decrease the integration time for a fixed observing hours per night to reduce this excess CV. However, one can always compute the total error variance starting from the scratch but it is not straightforward. We defer this to our future analysis with HERA. We would like to emphasize that this change is incorporated only for HERA, and not for the NenuFAR or SKA-Low.

Fig. \ref{fig:HERA_pred} show the predictions for observing the diagonal of 21-cm MAPS $\mathcal{D}(\nu,\nu)$ for four different $\ell$ bins in four panels. Every panel here is divided into two subpanels where, in the top subpanel we show the $\mathcal{D}(\nu,\nu)$ (thick blue lines) along with $2\sigma$ error estimates (\textit{i.e.} $2\boldsymbol{\sigma}_{11}^{\ell_{i}}$) for different observation time $t_{\rm obs}$, and in the bottom subpanel we plot SNR for the corresponding $t_{\rm obs}$ values. Here we have assumed that the foreground is perfectly modelled and completely removed from the observed data. However, this is not the case for HERA which avoides `Foreground wedge' in the ($\kk_{\perp}, k_\parallel$) plane \citep[e.g.][]{Datta_2010, Pober_2013} to mitigate foregorunds in power spectrum estimation. Therefore, the results presented below is optimistic and provide upper limits to the sensitivity. One can follow the method described in section 6 of \citet{Mondal_2020a} to incorporate the effect of foreground avoidance in an approximate way. However, we restrict ourselves to forground removal in order to make a fair comparison with NenuFAR. Note that we fix $2\sigma$ as the detection criterion in this paper. Considering any panel, the solid red and the solid orange lines represent the $2\sigma$ total error (top subpanel) and the corresponding SNR (bottom subpanel) respectively for $t_{\rm obs}=104~{\rm h}$ and $1000~{\rm h}$. We also plot the system noise contribution to the total error by red and orange dotted lines respectively for both $t_{\rm obs}=104\,{\rm h}$ and $1000~{\rm h}$ in the top subpanels. We also plot the $2\sigma$ CV (green dashed line) which is inherent to the signal and can be achieved after hypothetically integrating the system noise for $t_{\rm obs} \rightarrow \infty$. The dot-dashed grey line in each of the bottom subpanels demarcates the floor SNR$=2$ above which we claim it to be a detection.

Considering $\ell = 343$ (top left panel), we find that the 21-cm MAPS will be detectable (\textit{i.e.} SNR$\geq 2$) for $t_{\rm obs}=104~{\rm h}$ within frequency range $80-132~{\rm MHz}$ which contains the heating peak for our fiducial model. The SNR here can go as high as $4$ at $\nu \sim 110~{\rm MHz}$ which is closer to the CV limit. Increasing $t_{\rm obs}$ is expected to improves the SNR $\propto t_{\rm obs}$ for frequencies where the total error is dominated by system noise ($\nu>130~{\rm MHz}$). For $t_{\rm obs}=1000~{\rm h}$, we find that 21-cm MAPS at $\ell=343$ can be measured roughly across the full CD-EoR ($\nu >75 ~{\rm MHz}$). We also note that the system noise contribution $\sigma_{\rm N}$ increases rapidly towards the smaller frequency as the sky becomes brighter at large redshifts (see eq. \ref{eq:CN}). Considering the next bin at $\ell=538$ (top right panel), we find that the signal MAPS as well as its CV do not change much. However we find that $\sigma_{\rm N}$ relatively increases thereby increasing the total error in the measurement. This restricts the detection of the 21-cm MAPS to a narrower frequency range, \textit{i.e.} between $\nu =96-130~{\rm MHz}$ for $104~{\rm h}$ and $\nu \gtrsim 82~{\rm MHz}$ for $1000~{\rm h}$. We find that $\sigma_{\rm N}$ contribution to total error increases rapidly towards the larger $\ell$ values (small scales). This is due to the decrease of baseline distribution $\tau_{8}(\U_{{\rm g}_{i}})$ towards larger $U=\ell/(2\pi)$, and this decreases in baseline density is very rapid for compact arrays such as HERA. From the bottom panels of Fig. \ref{fig:HERA_pred}, we note that $104~{\rm h}$ of observations will be able to measure the 21-cm MAPS only within $\nu \approx 100-120~{\rm MHz}$ for $\ell =853$ where the CD MAPS peaks. For $t_{\rm obs}=1000~{\rm h}$ the $2\sigma$ detection is restricted only to limited frequency ranges between $87-132~{\rm MHz}$ for $\ell = 853$ and $100-125~{\rm MHz}$ for $\ell=1347$. The SNR drops considerably for any bins larger than $\ell=1347$ due to sudden fall in baseline sampling on the cells $\U_{\rm g}$.

%----------------------------------------------------------------------------------------------------------------------------------%

\subsection{Detectability with NenuFAR}\label{subsec:nenufar}
\begin{figure*}
\begin{center}
\includegraphics[scale=0.5]{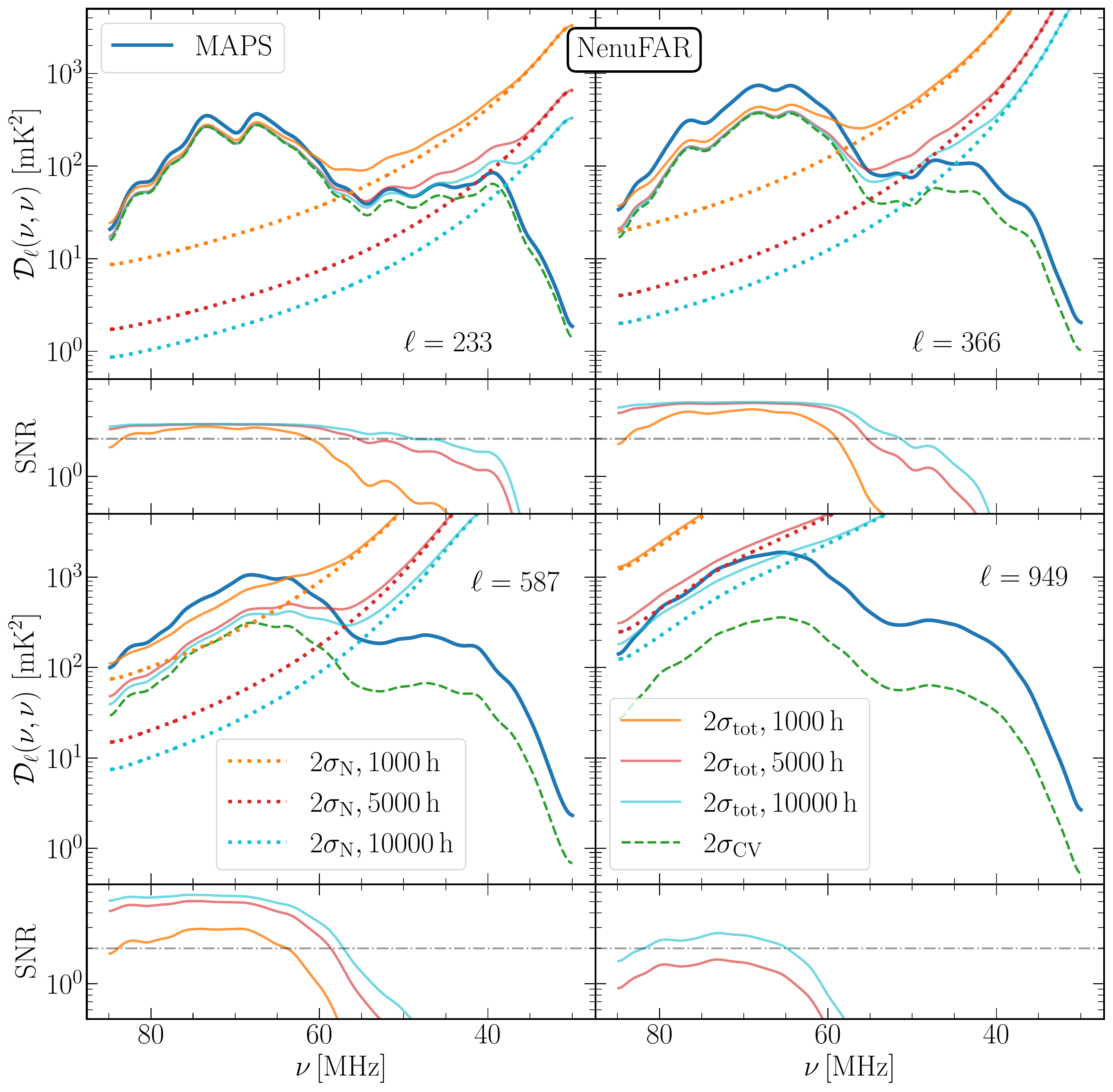}
    \caption{NenuFAR predictions for detecting the diagonal $(\nu_1=\nu_2)$ of the scaled MAPS $\mathcal{D}_{\ell}(\nu,\nu)=\ell (\ell+1) \mathcal{C}_\ell(\nu,\nu)/(2\pi)$. The four panels corresponds to the different $\ell$ bins as labelled. The thick solid blue lines show the expected CD-EoR 21-cm MAPS as a function of $\nu$ corresponding to M$3$-$f_{\rm X}10$. The red, orange and cyan solid lines in the four bigger panels represents the total $2\sigma$ error estimates, respectively for $1000\,{\rm h}$, $5000\,{\rm h}$ and $10000\,{\rm h}$ of NenuFAR observations. Corresponding dotted red, orange and cyan lines represents the system noise contribution ($2\sigma$) to the total error for $1000\,{\rm h}$, $5000\,{\rm h}$ and $10000\,{\rm h}$ respectively. The dashed green lines in the four bigger panels represents the respective CV error ($2\sigma$) estimates. The four smaller panels, corresponding to the bottom of each bigger panel, show how SNR of the measurements varies with the observing frequency $\nu$ for $1000\,{\rm h}$ (red), $5000\,{\rm h}$ (orange) and $10000\,{\rm h}$ (cyan). The gray dashed lines in the smaller panels demarcates SNR$=2$ floor for detection.}
   \label{fig:NF_pred}
\end{center}
\end{figure*}

NenuFAR$\textsuperscript{\ref{ft:nenu}}$ is an extension of the current LOFAR$^\textsuperscript{\ref{ft:lofar}}$ telescope that has been installed very recently at the Nan\c{c}ay Radio Observatory in France. It is expected to consist of $96$ core stations (a.k.a. mini-arrays) which are distributed within a compact circle of $\sim 400\,{\rm m}$ in diameter \citep{NenuFAR}. Apart from the compact core, it will have $6$ more mini-arrays as remote-stations. Currently, NenuFAR has $80$ core and $2$ remote stations rolled out and rest will be installed in future. Each mini-array consists of $19$ crossed dipole antennas arranged within a regular hexagonal pattern. NenuFAR operates within a frequency bandwidth of $10-85~{\rm MHz}$ that corresponds to a redshift range $z\approx 16-140$ spanning CD and Dark Ages. NenuFAR has already started its first phase of observations. We consider the channel width for NenuFAR to be twice of HERA, \textit{i.e.} $195.2~{\rm kHz}$. In this work, we consider a mock observation tracking a field at DEC$= +47^{\circ}22^{\prime}34^{\prime\prime}$ with the final phase of NenuFAR having $102$ mini-arrays. As NenuFAR operates in a very low frequency band, its $T_{\rm rec}$ is expected to be larger compared to the other telescopes considered here. For the purpose of this work we have assumed a mean value of $T_{\rm rec}=1000~{\rm K}$ \footnote{\href{https://nenupy.readthedocs.io/en/latest/instru/instrument_properties.html}{nenupy.readthedocs.io/en/latest/instru/instrument\_properties.html}} (see Table \ref{tab:params}) which remains smaller than $T_{\rm sky}$ at the relevant frequencies.

Note that our LC boxes span a frequency range between $\approx 70-200~{\rm MHz}$ where the X-ray heating, which is important during CD, becomes effective at $\nu > 80~{\rm MHz}$ for our fiducial model M$3$-$f_{\rm X}10$ (see Figs. \ref{fig:tb_slice_model} and \ref{fig:tb_slice_scenario}). This does not overlap at the low-frequency end with the observational frequency bandwidth of NenuFAR. Hence, we artificially shift our LC box by $40~{\rm MHz}$ towards smaller frequencies. The shifting in frequency, in principle, demands a change in $\delta_{\rm B}$ and therefore $\dtb(\mathbfit{x},\nu)$ due to evolution in background cosmology. However, for this analysis, we have assumed that the effect of the change in cosmology can possibly be overridden by properly tuning the physical processes by tweaking the relevant CD parameters within its uncertainties. We finally use the shifted LC box to make predictions for NenuFAR. The prediction here is to demonstrate the detectability of the NenuFAR in case the CD falls within its frequency bandwidth, and can not be directly compared with that of HERA. An better treatment of the signal is required for more accurate predictions and that we defer to our future work which will be devoted to NenuFAR.

As we have shifted the box in frequency, the extent of the accessible $\ell$ plane for NenuFAR will be a little different from HERA. Here the square $\ell$ plane has extent between $\pm 12093$ with $\Delta \ell = 108.95$, and we divide this also in $10$ log-spaced circular bins. Fig. \ref{fig:NF_pred} shows the predictions for measuring 21-cm MAPS using NenuFAR. Note that the following prediction considers an optimistic scenario where the foreground is assumed to be completely removed from the observed data. The plot is arranged in a similar way as in Fig. \ref{fig:HERA_pred}, except here we plot the $2\sigma$ errors and SNR for $1000~{\rm h},~5000~{\rm h}$ and $10000~{\rm h}$ in orange, red and cyan lines respectively. Our choices of $t_{\rm obs}$ here is not completely unrealistic as the data for more than $1000~{\rm h}$ of observations have already been acquired with NenuFAR. Considering the smallest bin at $\ell = 233$ (top left panel of Fig. \ref{fig:NF_pred}), we find that $1000~{\rm h}$ of NenuFAR observations can measure the X-ray heating peak of 21-cm MAPS only within $20~{\rm MHz}$ band around $\nu = 70~{\rm MHz}$ with SNR$\gtrsim 2$. Increasing $t_{\rm obs}$ to $5000$ or $10000~{\rm h}$ pushes the system noise contribution $\sigma_{\rm N}$ well below the CV error $\sigma_{\rm CV}$ at $\nu \gtrsim 50 ~{\rm MHz}$ and the total error $\sigma_{\rm tot}$ approaches the CV limit. We find that a detection between $55-85~{\rm MHz}$ is possible with $2< {\rm SNR} <3$. The SNR achieved for this bin is low due to small number of $\U_{\rm g}$ cells resulting into a larger CV contribution.

Moving to next bin, \textit{i.e.} $\ell = 366$ (top right panel), the relative contribution of CV to the total error decreases as expected. The system noise and the 21-cm MAPS is seen to increase in a similar proportion thereby improving the SNR as compared to $\ell =233$. We find that, for this bin and $t_{\rm obs} =1000~{\rm h}$, the 21-cm MAPS will be detected within a frequency range $58-84~{\rm MHz}$ with $2 \leq {\rm SNR} \leq 3$. This frequency range further expands to $\nu$ between $55-85~{\rm MHz}$ and $50-85~{\rm MHz}$ for $t_{\rm obs} \geq 5000~{\rm h}$ and $10000~{\rm h}$ respectively, with SNR becoming as large as $4$. The detection is still not possible for $\nu \lesssim 55~{\rm MHz}$ as the system noise increases rapidly with decreasing $\nu$ for larger $\ell$ values due to a decrease in the baseline density distribution.   

The system noise contribution increases further for $\ell = 587$ (bottom left panel), where we note that $t_{\rm obs} =1000~{\rm h}$ is able to measure the peak of the MAPS within $64-84~{\rm MHz}$ at $2 \leq {\rm SNR} \leq 3$. For $t_{\rm obs} = 5000$, the frequency range of the detected MAPS increases to $58-85~{\rm MHz}$ with a maximum SNR$\approx 5$. This further improves for $t_{\rm obs} =10000~{\rm h}$, however, not significantly. As is apparent from Fig. \ref{fig:NF_pred}, the CV contribution $\sigma_{\rm CV}$ decreases towards larger $\ell$ values (bottom panels) as there are more cells. However, the contribution from $\sigma_{\rm N}$ increases at a faster rate with increasing $\ell$. This is due to the fact that the baseline density distribution decreases rapidly at large $uv$ distances because NenuFAR has fewer correlated receiver elements (mini-arrays) than HERA. Considering  $\ell = 949$ (bottom right panel), which is almost dominated by system noise, we note that $t_{\rm obs} \gtrsim 10000~{\rm h}$ of observations is required to measure the signal MAPS at SNR$\geq 2$ and that still within a narrow frequency band around $\nu = 70 ~{\rm MHz}$. The system noise increases very rapidly for the bins with $\ell \gtrsim 1000$ for NenuFAR making it impractical to measure the signal MAPS at small-angular scales.

%----------------------------------------------------------------------------------------------------------------------------------%

\subsection{Detectability with SKA-Low}\label{subsec:ska}
\begin{figure*}
\begin{center}
\includegraphics[scale=0.5]{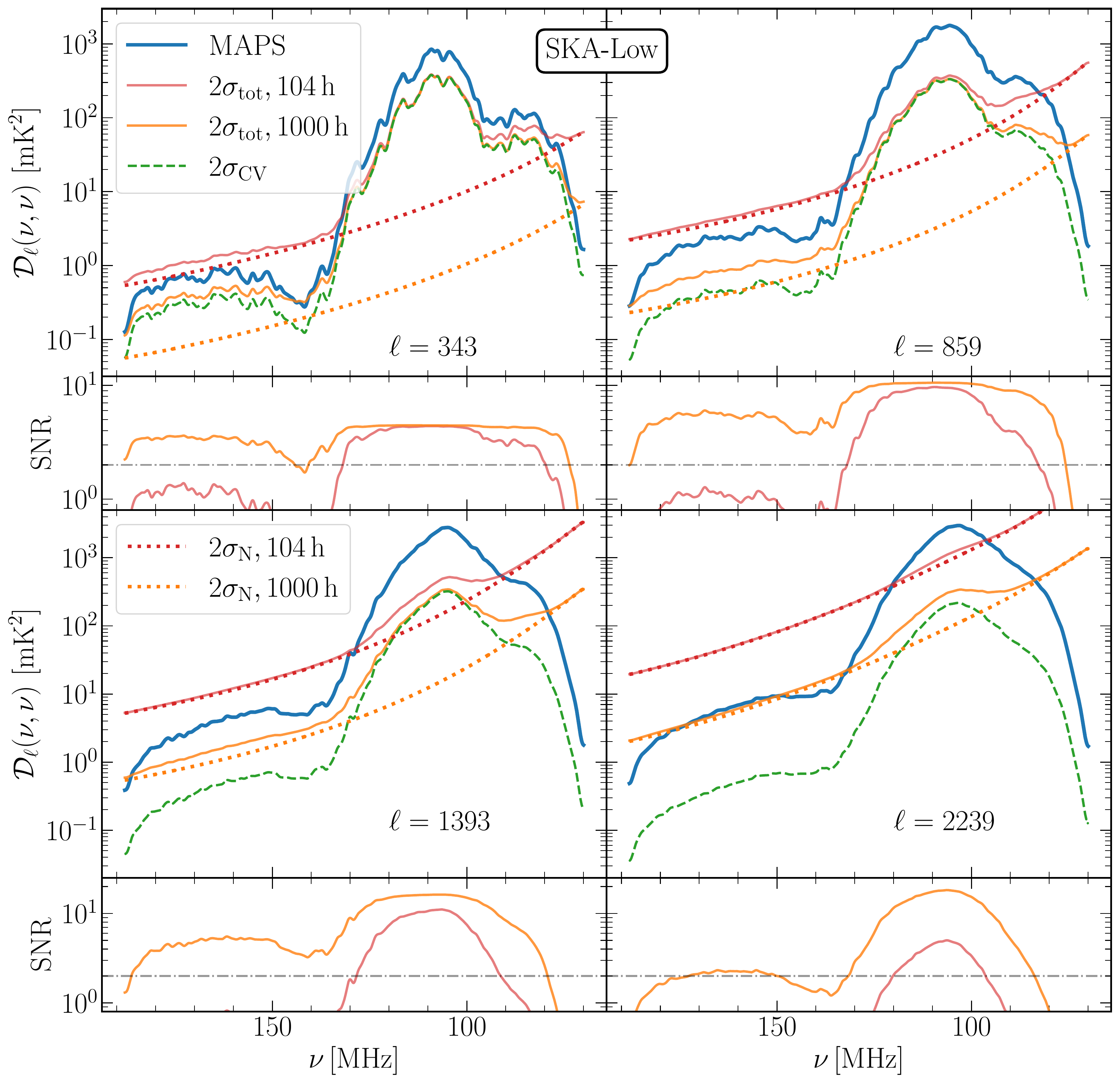}
    \caption{SKA-Low predictions for detecting the diagonal $(\nu_1=\nu_2)$ of the scaled MAPS $\mathcal{D}_{\ell}(\nu,\nu)=\ell (\ell+1) \mathcal{C}_\ell(\nu,\nu)/(2\pi)$. The arrangement of this plot is the same as in Fig. \ref{fig:HERA_pred}.}
   \label{fig:SKA_pred}
\end{center}
\end{figure*}

The Square Kilometer Array (SKA$^\textsuperscript{\ref{ft:ska}}$) is going to be world's largest radio telescope which is expected to be completed by the end of this decade. It consists of two arrays -- (1) SKA-Mid, which is being built at the Karoo desert in South Africa and will operate within the frequency range of $350 - 15300~{\rm MHz}$ and (2) SKA-Low, which is being built at Murchison Radioastronomy Observatory in Western Australia and will operate within the frequency range of $50-350~{\rm MHz}$. SKA-Low will be suitable for the CD-EoR measurements \citep[][]{koopmans2014} as it can observe 21-cm signal coming from the redshifts between $z= 3$ and $27$. Here we consider the channel width of SKA-Low to be the same as HERA, \textit{i.e.} $97.66~{\rm kHz}$. SKA-Low will have $\sim 131$ thousands log-periodic dipoles which will be distributed among $512$ stations in its final phase. Each of the stations will have $256$ dipoles that will be scattered within a diameter of $35~{\rm m}$. The antenna layout of SKA-Low will have a compact core with roughly $50~\%$ of stations lying within a diameter of $\sim 1 ~{\rm km}$. The remaining stations will be grouped into clusters of $6$ stations and these clusters will be arranged in three modified spiral arms such that the maximum baseline length will be $\sim 64 ~{\rm km}$ \citep{SKA_Low_v2}. Here we consider an observation with SKA-Low tracking a field at DEC$=-26^{\circ}49^{\prime} 29^{\prime\prime}$ for $8$~h/night. We use the resulting baseline tracks to predict the errors in measuring MAPS with this telescope. For SKA-Low, we have used similar $\ell$ plane and binned it on the same way as for HERA.

Fig. \ref{fig:SKA_pred} shows the errors estimated in the measurements of $\mathcal{D}_{\ell}(\nu,\nu)$ for $t_{\rm obs}=104~{\rm h}$ and $1000~{\rm h}$, just as we have presented for HERA but for $\ell = 343,~859,~1393$ and $2239$. We note that, the predictions here also consider foreground removal for comparison on similar grounds. The color scheme of this figure is exactly the same as that in Fig. \ref{fig:HERA_pred}. Comparing SKA-Low with the previous telescopes, it is able to perform far better in terms of sensitivity (Figs. \ref{fig:HERA_pred}, \ref{fig:NF_pred} and \ref{fig:SKA_pred}). Also having a denser baseline distribution and longer $uv$ extent allows SKA-Low to measure MAPS on smaller scales ($\ell \approx 5000$) compared to compact arrays such as HERA and NenuFAR. Considering the four panels of Fig. \ref{fig:SKA_pred}, we note that SKA-Low will be able to measure the CD 21-cm MAPS with $104~{\rm h}$ of observations. We find that a $2\sigma$ measurement is possible within the frequency ranges $80-132~{\rm MHz},~ 82-132~{\rm MHz},~ 90-128~{\rm MHz}$ and $96-120~{\rm MHz}$ respectively for $\ell = 343,~859,~1393$ and $2239$. We find that the system noise contribution to the total error $\sigma_{\rm tot}$ drops considerably for $t_{\rm obs}=1000~{\rm h}$ allowing SNR to rise above the detection limit of $2$ in almost the whole frequency bandwidth for the four $\ell$ bins considered here. However the SNR does not improve much around the X-ray heating peak in 21-cm MAPS ($\nu \approx 100~{\rm MHz}$) by increasing $t_{\rm obs}$ beyond $104~{\rm h}$ for $\ell \lesssim 1000$ as the $\sigma_{\rm tot}$ is dominated by $\sigma_{\rm CV}$. In such cases, it is better to divide the total observation hours over multiple independent sky-patches. This can be done for the field-tracking instruments such as NenuFAR and SKA-Low. Combining the data incoherently from different fields will help in beating down the CV to the measured MAPS \citep[e.g.][]{koopmans2014}.

Considering top panels for $t_{\rm obs}=1000~{\rm h}$, we find that the measurement of 21-cm MAPS is possible for $\nu \geq 75~{\rm MHz}$ with maximum SNR$\approx 4$ and $10$, respectively for $\ell=343$ and $859$. However, we note that the detection is possible at $\nu \geq 80~{\rm MHz}$ for $\ell=1393$ and between $85-170~{\rm MHz}$ for $\ell=2239$. The maximum value of SNR saturates to values near $15$ where $\sigma_{\rm tot}$ is dominated by CV for $\ell = 1393$. For $\ell=2239$ the SNR rises further to a maximum value of $20$ near the heating peak ($\nu \approx 105~{\rm MHz}$) of the 21-cm MAPS. The system noise contributions $\sigma_{\rm N}$ to $\sigma_{\rm tot}$ gradually increases in the larger $\ell$ bins as the baseline density decreases. This further starts decreasing the SNR, and it requires larger values of  $t_{\rm obs}$ to achieve a sufficient SNR to detect the signal from small angular-scales (large $\ell$).

\begin{table}
    \centering
    \tabcolsep 10pt
    \renewcommand\arraystretch{1.5}
    \begin{tabular}{c c c c}
    \hline
    Telescope & ${\rm SNR_{max}}$ & Peak freq. & ${\rm SNR_{int}}$ \\
     & (per channel) & ($\nu_{\rm p}~ {\rm in~ MHz}$) & (${5~\rm MHz}$) \\
    \hline
    \hline
    HERA & $5.9$ & $105$ & $6.2$ \\
    NenuFAR & $3.0$ & $67.5$ & $5.5$ \\
    SKA-Low & $6.4$ & $105$ & $6.5$ \\
    \hline
    \end{tabular}
    \caption{Shows the maximum SNR achievable (${\rm SNR_{max}}$) per frequency channel for the three instruments at $\ell \sim 550$ with $1000~{\rm h}$ of observations. We also show ${\rm SNR_{int}}$ integrated within $5~{\rm MHz}$ frequency-width around frequencies at which $\mathcal{D}_{\ell}(\nu,\nu)$ peaks, \textit{i.e.} $\nu_{\rm p}$.}
    \label{tab:snr}
\end{table}
%%%%%%%%%%%%%%%%%%%%%%%%%%%%%%%%%%%%%%%%%%%%%%%%%%%%%%%%%%%%%%%%%%%%%%%%%%

\section{Summary and Discussions}
\label{sec:summ}

The measurements of brightness temperature fluctuations $\delta T_{\rm b}$ of the redshifted 21-cm radiation from the \HI~provides a unique way to unveil the secrets of formation of the first luminous objects in the Universe. The primary goal of all existing and future radio-interferometric experiments (e.g. LOFAR, MWA, HERA, NenuFAR, uGMRT, SKA-Low) is to measure the 3D power spectrum $\Delta^2_{\rm b}(k,z)$ of the fluctuating 21-cm signal coming from the Cosmic Dawn (CD) and Epoch of Reionization (EoR). The definition of 3D power spectrum inherently assumes the signal to be statistically homogeneous and isotropic along all directions of the signal volume. However the CD-EoR 21-cm signal in real observations is not statistically homogeneous and isotropic along the line of sight (LoS) direction of the observed volume due to light-cone (LC) effect \citep[][]{Barkana_2006}. The LC effect on the $\Delta^2_{\rm b}(k,z)$ is considerable for EoR 21-cm signal at least on large-scales \citep[e.g.][]{Light_cone_I, Zawada_2014, Mondal_2017}. This effect on 3D power spectrum becomes considerable during CD when the 21-cm signal is controlled by the spin-temperature fluctuations driven by the X-ray sources \citep{Ghara_2015b}. Also, we find that the deviations in 3D power spectrum due to LC effects during CD can be as large as $\pm 100\%$ for our fiducial model (M$3$-$f_{\rm X}10$) at scales corresponding to $k=0.15~{\rm Mpc}^{-1}$ (see Fig. \ref{fig:3dps}). The deviations are found to decrease on small scales ($k=1.1~{\rm Mpc}^{-1}$), however they are still closer to $\pm 100\%$ for our fiducial model. Note that, here we consider a large comoving volume $V=[500h^{-1}~{\rm Mpc}]^3$ which corresponds to a frequency bandwidth $\sim 40 ~{\rm MHz}$ at relevant frequencies which are not very unrealistic for the HERA and SKA-Low. However the LC effects will be less pronounced if the data is analyzed within a smaller frequency band, say $\sim 10~{\rm MHz}$, as is being done for LOFAR and NenuFAR etc.

The Multi-frequency Angular Power Spectrum (MAPS) (eq. \ref{eq:cl}) provides an unbiased alternative to the 3D power spectrum. In this work, we use the bin-averaged MAPS \citep{Mondal_2017} estimator (eq. \ref{eq:bin-cl}) to quantify the two-point statistics of the simulated (see section \ref{sec:sim}) LC of the CD-EoR 21-cm signal. We find that the diagonal 21-cm MAPS $\mathcal{D}_{\ell}(\nu,\nu)$ captures the signatures of cosmic evolution of the signal more distinctly than the 3DPS (see Figs. \ref{fig:3dps}, \ref{fig:MAPS_diag_models} and \ref{fig:MAPS_diag_scenario}) when analyzed within a large-bandwidth signal volume. However, this is a matter of binning and analyzing the data. It is still possible to see the distinct evolutionary features in 3DPS by computing it from the data divided into small-bandwidth chunks, within which the signal is considerably correlated (see Fig. \ref{fig:MAPS_corr}). Analyzing data in small bandwidths reduces the LC bias in the 3DPS estimates, if not removed. The shape and the amplitude of the 21-cm MAPS are sensitive to the properties of the sources and the processes driving CD-EoR. This extra information has been tentetaively shown to provide more stringent constraints over the EoR model parameters than the 3DPS \citep[][]{Mondal_2022}. In a future work, we plan to explore the impact of MAPS on the CD model parameter estimation, relative to the 3DPS.

Considering a model where the spin temperature is saturated and $T_{\rm s} >> \TCMB$ (here M$1$), we only see a dip around a point where the reionization starts and a peak when IGM is $\sim 50\%$ ionized. This feature in the MAPS of model M$1$ is distinctly different from the other models (see Table \ref{tab:sim}) where X-ray heating is included in the simulations self-consistently from the beginning. We also note that the 21-cm MAPS of M$1$ is roughly two orders of magnitude smaller than that of the other models with self-consistent evolution of $T_{\rm s}$. In the other two models M$2$ and M$3$, the unsaturated X-ray heating boosts 21-cm MAPS during CD by introducing additional fluctuations in $\dtb$ through $T_{\rm s}$. We distinctively note a peak like feature around $\nu=110~{\rm MHz}$ in the diagonal and near-diagonal elements (see Figs. \ref{fig:MAPS_full_models} and \ref{fig:MAPS_diag_models}) for models M$2$-$f_{\rm X}10$ and M$3$-$f_{\rm X}10$. The location of the heating peak and its width depends upon the X-ray efficiency $f_{\rm X}$ of the sources in the IGM as seen in Figs. \ref{fig:MAPS_full_scenario} and \ref{fig:MAPS_diag_scenario}. We also note another peak in 21-cm MAPS around $\nu = 90 ~{\rm MHz}$ which arises due to fluctuations in the Ly-$\alpha$ coupling in the model M$3$-$f_{\rm X}10$ (see Figs. \ref{fig:MAPS_full_scenario} and \ref{fig:MAPS_diag_scenario}). The peak is prominent for large angular-scales (small $\ell$) and it gradually merges with the heating peak and disappears as we move to the large $\ell$ values.

We also find the correlations between the signal at different frequencies ($\nu_1,\nu_2$) significantly drops roughly beyond $\Delta \nu = \pm 3~{\rm MHz}$ (Fig. \ref{fig:MAPS_corr}). The rate of decorrelation depends upon the concerned epoch and also it can be related to the mean free path of the photons which plays important role in governing the signal during that epoch. We note that the epoch where X-ray heating plays important role ($\nu \approx 100-140~{\rm MHz}$), the signal remains positively correlated over a larger frequency range (see Figs. \ref{fig:MAPS_full_models}, \ref{fig:MAPS_offdiag_ratio} and  \ref{fig:MAPS_full_scenario}) as the X-rays can travel farther. On the other hand, the decorrelation during EoR is faster as the mean free path of the UV photons is much less than that of the X-rays.

The observed 21-cm signal from CD-EoR gets contaminated with the system noise, foregrounds, RFI and other systematics. This contributes to the error in the measurement of 21-cm MAPS along with the inherent cosmic variance (CV). In this work, we present the detectability of the 21-cm MAPS $\mathcal{D}(\nu,\nu)$ in the context of future observation using HERA, NenuFAR and SKA-Low. We estimate the expected errors in the measured 21-cm MAPS considering the observed signal to be free from systematics, foregrounds and RFI. We note that NenuFAR follows the foreground removal strategy, whereas HERA is designed to avoid the foreground wedge in their power spectrum analysis. However for an equal comparison, we assume that foreground is perfectly subtracted from the data, for every telescope considered here, which leads to optimistic sensitivity predictions. Here we assume that the 21-cm signal is statistically equivalent to a Gaussian random field and ignore the impact of trispectrum in the CV measurements \citep[e.g.][]{Mondal_I}. The contribution from non-Gaussianity to the error estimates can be important only during later stages of EoR where the trispectrum becomes comparable or stronger than the Gaussian system noise contribution \citep[][]{shaw_2019}. During CD, the Gaussian system noise will be much larger than the inherent non-Gaussianity of the signal. 

We find that $104~{\rm h}$ of HERA drift scan observations will be able to provide the CD 21-cm MAPS (diagonal elements) at $> 2\sigma$ level for $\ell \lesssim 1000$ with the errors being dominated by CV below $\ell \approx 400$ (see Fig. \ref{fig:HERA_pred}). With $1000~{\rm h}$ of observations, we can achieve $>2\sigma$ measurement of CD 21-cm MAPS for $\ell \lesssim 1800$. HERA is able to measure the EoR  21-cm MAPS only for $\ell < 850$ with $1000~{\rm h}$ observations. We can obtain the maximum SNR of $\approx 8$ with $1000~{\rm h}$ of data for the bin corresponding to $\ell=853$ (see Fig. \ref{fig:HERA_pred}). Increasing the observation hours further can increase the SNR up to $10$ which is fixed by the CV for this bin. The sensitivity of HERA decreases for larger $\ell$ bins due to compact baseline distribution and we can not measure signal beyond $\ell = 1800$.

The upcoming SKA-Low is expected to outperform HERA in terms of sensitivity due to denser baseline densities. The system noise contribution is much less for SKA-Low that the total error in the measurement of CD 21-cm MAPS is dominated mostly by the CV contribution for $\ell \lesssim 1000$ with $t_{\rm obs} = 104~{\rm h}$. Increasing $t_{\rm obs}$ further to $1000~{\rm h}$ does not improve the SNR much for the CD signal. However the detectability improves considerably for the EoR 21-cm MAPS and it reaches ${\rm SNR} > 2$ for $t_{\rm obs}=1000~{\rm h}$. The system noise increases and becomes comparable with the CV contribution for $\ell > 1000$, however it is still possible to measure the peak of the CD 21-cm MAPS with $t_{\rm obs} = 104~{\rm h}$ for $\ell < 5000$. The EoR 21-cm MAPS can only be detected at ${\rm SNR}>2$ within $\ell < 2240$, but for $t_{\rm obs} = 1000~{\rm h}$. The SNR for SKA-Low can be as high as $10$ and $20$ respectively for $t_{\rm obs} =104~{\rm h}$ and $1000~{\rm h}$ even at intermediate scales of $\ell \sim 1000$ (see Fig. \ref{fig:SKA_pred}). 

NenuFAR operates in a frequency band $10-85~{\rm MHz}$ which probes the 21-cm signal farther in redshift than HERA and SKA-Low. It has channels which are twice as wide as for HERA and SKA-Low. Given this mismatch in the frequency band with the other telescopes, we consider a different CD-EoR model for NenuFAR predictions. This new model assumes that heating started at earlier redshifts such that the strong impact of spin temperature fluctuations falls within the frequency band of NenuFAR. If the peak of CD 21-cm MAPS falls in the range of NenuFAR, it will be possible to measure it at ${\rm SNR}>2$ for $\ell \lesssim 600$ with $t_{\rm obs}=1000~{\rm h}$ (see Fig. \ref{fig:NF_pred}). The SNR will improve if $t_{\rm obs}$ is increased to say $5000~{\rm h}$ or $10000~{\rm h}$ but the noticeable change is only restricted to a narrow range of $360 < \ell < 600$. The reason for such large error in the measurement is the system noise contribution which increases very rapidly towards large redshifts. It is not possible to achieve an ${\rm SNR} \geq 2$ per frequency channel even with $t_{\rm obs}=10000~{\rm h}$ beyond $\ell = 1000$. 

The whole point of computing MAPS by correlating the 21-cm signal across the frequency channel elements is for keeping the LC evolution information from the signal volume. However, analyzing the data over small frequency channels makes the 21-cm MAPS statistics suffer from the noise contribution. Whereas in 3DPS, we analyze the data over a large frequency bandwidth that effectively reduces the noise contribution in 3DPS estimates. Contradicting the philosophy behind MAPS, one can anyway improve the SNR of the MAPS measurements by combining the observed signal in consecutive frequency channels. This way of beating down noise is effective only when the measurement error is dominated by the system noise such as in the case of NenuFAR. We find that averaging the signal within a frequency band of $5~{\rm MHz}$ around the CD MAPS peak frequency considerably improves the integrated SNR for NenuFAR (see the last column of Table \ref{tab:snr}). However there is small change for HERA and negligible change for SKA-Low as compared to the SNR per channel estimated at $\ell \sim 550$ for $t_{\rm obs} =1000~{\rm h}$. This is because the error is CV dominated for both HERA and SKA-Low.

It is possible to further reduce the CV and improve SNR for field-tracking instruments (e.g. SKA-Low, NenuFAR) by breaking up one long observation (say $\geq 1000~{\rm h}$) to shorter observations targeting different sky-patches and combining them afterwards. However, the division should be such that the system noise SNR computed from an individual patch remains well above unity \citep[e.g.][]{koopmans2014}. Nonetheless the channel averaging can still be useful to achieve a decent integrated SNR for HERA and SKA-Low at large $\ell$ and/or for smaller $t_{\rm obs}$ where the system noise remains dominant. However one needs to choose the frequency range carefully such that the 21-cm signal MAPS does not decrease too much especially near the peaks.

Our study finally concludes that it will be possible to measure the CD-EoR 21-cm MAPS within reasonable observation time using the current and upcoming radio-interferometers mentioned in this paper. 
%The predictions presented here is subjected to a foreground removal case. 
Incorporating the effects of foregrounds, however, can degrade the prospect of detection \citep[][]{Mondal_2020a}. We also ignore the effects of non-Gaussianity of the 21-cm signal which can affect the error predictions and hence the detectability, especially during the mid and later stages of EoR. It is therefore important to include the impact of foreground as well as that of the inherent non-Gaussianity of the signal to make realistic predictions, and we defer these tasks for our future work.
%--------------------------------------------------------------------

\section*{Acknowledgements}
AKS, RG and SZ acknowledge support by the Israel Science Foundation (grant no. 255/18). RM is supported by the Israel Academy of Sciences and Humanities \& Council for Higher Education Excellence Fellowship Program for International Postdoctoral Researchers. GM acknowledges support by Swedish Research Council grant 2020-04691. FM acknowledges support of the PSL Fellowship. LVEK acknowledges the financial support from the European Research Council (ERC) under the European Union's Horizon 2020 research and innovation programme (Grant agreement No. 884760,``CoDEX"). The authors thank the anonymous referee for the constructive feedback.
%--------------------------------------------------------------------

\section*{Data Availability}
The data underlying this article will be shared on reasonable request to the corresponding author.

%--------------------------------------------------------------------

\bibliography{ref}

%--------------------------------------------------------------------
%\appendix
%\section{}
%\label{a1}
%--------------------------------------------------------------------

%\vfill
\bsp

\label{lastpage}

\end{document}